\documentclass[10pt,preprint,a4paper,aps,pra,showpacs,floatfix,raggedbottom,floats,superscriptaddress]{revtex4-2}

\usepackage{graphicx,latexsym}
\usepackage{amssymb,amsmath,bm}
\usepackage{mathrsfs}
\usepackage[T1]{fontenc} 
\usepackage[utf8]{inputenc}
\usepackage{xcolor}
\usepackage{ulem}
\DeclareMathOperator{\sech}{sech}

\begin{document}

\title{Suppression of Bloch Oscillations and Nonreciprocal Landau-Zener Tunneling in Bose-Einstein Quantum Droplets}

\author{Szu-Cheng Cheng}
\affiliation{Department of Optoelectric Physics, Chinese Culture University, Yang-Ming-Shan, Taipei 11114, Taiwan.}
\author{Yu-Wen Wang}
\affiliation{Graduate Institute of Photonics and Optoelectronics, National Taiwan University, Taipei 10617, Taiwan.}
\affiliation{Department of Applied Physics and Chemistry, University of Taipei, Taipei 10048, Taiwan.}
\author{Wen-Feng Hsieh}
\affiliation{Department of Photonics, National Yang Ming Chiao Tung University, Hsinchu 30010, Taiwan.}
\author{Vidar Gudmundsson}
\affiliation{Science Institute, University of Iceland, IS-107 Reykjavik, Iceland}
\author{Wen-Hsuan Kuan}
\email{wenhsuan.kuan@gmail.com}
\affiliation{Department of Applied Physics and Chemistry, University of Taipei, Taipei 10048, Taiwan.}
\affiliation{Science Institute, University of Iceland, IS-107 Reykjavik, Iceland}









        

%

\begin{abstract}
We investigate the nonlinear Bloch dynamics and Landau-Zener (LZ) tunneling of quantum droplets in optical lattices. We show that the Lee-Huang-Yang (LHY) correction not only stabilizes the self-bound droplet, but also introduces nonlinear phase feedback that competes with the lattice-induced coherent motion. In the deep-lattice regime, applying a generalized super-Gaussian ansatz within the tight-binding model demonstrates that chirp accumulation modifies the internal phase profile and renormalizes mobility. The coherent Bloch oscillations (BO) are progressively arrested in the presence of the LHY interaction without dissipative damping. 
In the shallow-lattice regime, the system is mapped onto a nonlinear two-level Josephson-analog model in which the mean-field and LHY contributions enter through an effective nonlinear detuning, deforming the adiabatic spectrum and generating looped bands. Using the classical action-angle formulation, we demonstrate that the nonlinear LZ tunneling is governed by the underlying phase-space structure. In particular, the LHY correction suppresses the tunneling probability by modifying the separatrix action and renormalizing the exponential sweep-rate scaling through a nonlinear weighting factor. We further identify pronounced nonreciprocal LZ tunneling arising from branch-dependent population imbalance and the nonlinearly induced inertia. These results establish a unified mechanism in which the LHY interaction suppresses both coherent Bloch dynamics and interband tunneling by reorganizing the dynamical exchange among lattice motion, population imbalance, and internal phase modulation.   
\end{abstract}

\maketitle

\section{Introduction}

Optical lattices have emerged as a versatile platform for studying quantum transport, enabling the realization of a broad range of many-body and single-particle phenomena in engineered periodic potentials. These include superfluid-Mott insulator transitions, Josephson dynamics, lattice interferometry, and topological band engineering, which span strongly correlated phases, coherent transport, and precision measurements \cite{Bloch2008, Lewenstein2017, Goldman2014, Cataliotti2001, Greiner2002, Viebahn2019}. The high degree of tunability in lattice geometry, dimensionality, and interactions has established optical lattices as a unique setting in which both equilibrium states and nonequilibrium dynamics can be explored and systematically controlled \cite{Price2017, Li2016}.

Within this broad landscape, BO and LZ tunneling provide a particularly direct window to the dynamics of driven lattice systems \cite{Morsch2001, Cristiani2002, Gustavsson2008, Natale2022}. In contrast to bulk solid-state materials, where BO is typically suppressed by scattering and dephasing, using optical lattices allow us to directly observe the long-lived coherent band dynamics. Beyond their roles as fundamental transport phenomena, BO serves as a sensitive probe of band structure and coherence, for applications ranging from lattice spectroscopy to precision force measurements. In parallel, LZ tunneling constitutes a paradigmatic mechanism for nonadiabatic transitions, enabling controlled population transfer and providing direct access to energy gaps and dynamical phase information. These processes establish a fundamental framework for exploring the coherent control of quantum transport in driven periodic systems.

A particularly rich setting for exploring interaction-driven dynamics is provided by quantum droplets stabilized by LHY corrections \cite{LHY1957, Petrov2015}. These self-bound states arise from the competition between the attractive mean-field interactions and the repulsive quantum fluctuations, leading to liquid-like phases in dilute ultracold gases. Such droplets have been experimentally realized in both binary Bose mixtures and dipolar condensates \cite{Cabrera2018, Cheiney2018, Semeghini2018, Barbut2016, Schmitt2016, Chomaz2016}, and exhibit a variety of collective phenomena including anisotropic deformation, surface excitations, and vortex formation \cite{Wenzel2017, Barbut2018, Tanzi2019, Cheng2024}. The underlying stabilization mechanism relies on nonlinear energy contributions that scale explicitly with dimensionality \cite{Petrov2016}. In three-dimensional dilute atoms, the mean-field and LHY terms scale as $n^2$ and $n^{5/2}$, respectively. In two-dimensional binary mixtures, the correction is modified to $n^2 \ln n$, while in one dimension it reduces to a balance between $n^2$ and $n^{3/2}$. This dimensional dependence represents a definite feature of beyond-mean-field quantum many-body physics.

Despite this intricate equilibrium and dynamical behavior, most studies of quantum droplets have focused on stationary or weakly driven regimes. Their transport properties in strongly modulated or accelerated optical lattices remain far less explored. Recent works have begun to address droplet dynamics in lattice environments, including survival and modulation in shallow lattices, excitation and oscillation induced by lattice shifts, and direction-dependent transport across periodic potentials \cite{Gao2025, Nie2023, Kartashov2024}. However, how the beyond-mean-field interactions modify the nonadiabatic tunneling processes in such driven settings remains an open question.

In this work, we address how the LHY corrections reshape BO and LZ tunneling in driven optical lattices. We show that the droplet dynamics can be consistently described across deep- and shallow-lattice regimes within a unified framework. In the deep-lattice limit, we employ a tight-binding description combined with a generalized super-Gaussian variational ansatz, which can capture the flat-top or soliton-like density profile characteristic of droplets and provides a more accurate representation than the conventional Gaussian approximations. The generation of these beyond-Gaussian structures within nonlinear media or technologies is increasingly relevant for quantum control and information processing in ultracold, telecommunication, and ultrafast laser systems \cite{chun-chia2022, Clark2016, Derevyanko2016, Blanco-Redondo2016, kuan2018}. Within this framework, we further establish the equivalence between different implementations of lattice acceleration via a frame transformation, linking the real-space tilting to the momentum-space detuning.

Our analysis reveals that the LHY correction plays a crucial dynamical role beyond a simple renormalization of interaction strength. In the deep-lattice regime, it stabilizes the droplet against dispersion while introducing strong nonlinear feedback through chirp-mediated parametric couplings that suppress large-amplitude coherent motion under external driving. As a result, the oscillatory dynamics observed in the regime where the mean-field effect dominates is rapidly damped, and the system evolves toward dynamically stabilized or self-trapped configurations, rather than sustaining long-lived BO. In the shallow-lattice regime, the system can be reduced to a nonlinear two-level Josephson-analog model near the Brillouin zone edge \cite{Wu2003}, where interactions deform the adiabatic spectrum and generate looped band structures \cite{Liu2002}. By analyzing the associated phase-space dynamics using a classical action-angle formulation \cite{Landau1976, Goldstein2002}, we show that the tunneling processes are governed by two distinct mechanisms: an adiabatic regime controlled by the phase-space area enclosed by separatrix trajectories, and a nonadiabatic regime determined by the complex-phase structure of the action \cite{Bayfield1999}.

A central finding of the shallow-lattice studies is that the LHY corrections qualitatively suppress LZ tunneling by reducing the effective nonlinear detuning and modifying the phase-space topology. This leads to a systematic reduction in tunneling probability and a renormalization of the exponential scaling with the lattice acceleration. Furthermore, we identify a pronounced nonreciprocal tunneling behavior, where transitions from the lower to the upper bands differ from their reverse process under identical driving protocols. This asymmetry originates from the nonlinear population imbalance and its impact on the effective detuning, which drives the nonlinearly induced inertia and fosters a localization tendency governed by the LHY field.

These results establish a unified picture in which nonlinear band deformation, phase-space topology, and tunneling dynamics are intrinsically linked in driven quantum droplets. In particular, the LHY-induced suppression of both coherent Bloch dynamics and interband tunneling provides a common mechanism through which the beyond-mean-field interactions dictate transport in driven lattices. Our work demonstrates how the density and phase modulations of the nonlinear effects govern both adiabatic and nonadiabatic transport, and provides a framework for engineering interaction-controlled tunneling phenomena in ultracold atomic systems.

%
%

\section{Deep Lattices}

\subsection{Frame Transformation and Perspective Equivalence}
While BO requires an external driving force, the description of the lattice dynamics must incorporate acceleration. In optical-lattice experiments with neutral atoms, such an effective acceleration can be implemented by chirping the frequency difference between two counter-propagating laser beams, which produces a moving interference pattern. In the laboratory frame, the resulting lattice potential takes the time-dependent displaced form
\begin{align}
 V_{\rm lab}(x,t) = V_L(x-\alpha_L t^2/2),
\end{align}
where $\alpha_L$ denotes the lattice acceleration generated by the frequency chirp. 
However, BO probes the intraband dynamics and phase accumulation within a single Bloch band, for which an explicitly time-dependent lattice potential obscures the essential physics. It is therefore more natural to adopt a representation in which the lattice potential is static and the driving enters as an effective linear force, 
\begin{align}
V_{\rm ext} = V_L + F x. 
\end{align}
Although this tilted-lattice description is widely employed in theoretical studies of BO, its relation to the experimentally realized moving-lattice picture is not always clearly articulated.
In the following, we formally establish this correspondence by constructing the unitary transformation connecting these two representations, thereby clarifying that the static tilted lattice arises from a change of representation rather than from a physical translation of the lattice itself.

To this end, we consider the acceleration transformation
\begin{align}
x \rightarrow x' = x - \alpha_L t^2/{2}.
\end{align}
At the classical level, this transformation is formulated as a canonical transformation connecting the Hamiltonians $H(x,p)$ and $H'(x',p')$ in the laboratory and accelerating frames, differing only by a partial time derivative of the generating function. Following the classical--quantum correspondence established by Klink, this canonical transformation is promoted to a unitary transformation in quantum mechanics \cite{Klink1997}. We denote the corresponding acceleration operator by
\begin{align}
\hat{U} = e^{i p\, \alpha_L t^2/2\hbar},
\end{align}
which establishes the equivalence between the wavefunctions in the laboratory and accelerating frames, $\Psi'(x,t)=\hat{U}\Psi(x,t)$.
%
Applying the unitary operator $\hat U$ to the time-dependent Schr\"{o}dinger equation yields the Hamiltonian in the accelerating frame,
\begin{align}
\hat H' = \hat U \hat H \hat U^{-1} + i\hbar\,\dot{\hat U}\hat U^{-1}.
\end{align}
After combining the transformed kinetic energy with the inertial contribution arising from the additional term $i\hbar\,\dot{\hat U}\hat U^{-1}$, and neglecting the resulting purely time-dependent phases, the remaining position-dependent term gives the fictitious linear potential, 
\begin{align}
V_f = M \alpha_L x \equiv Fx,
\end{align}
where \(M\) is the atomic mass and \(F = M\alpha_L\) is the fictitious force generated by the lattice acceleration. This result reveals the gauge correspondence between the moving-lattice and the static tilted-lattice descriptions, which represent the same physical situation viewed in different frames.
An analogous equivalence underlies acceleration induced by a gravitational field in vertically oriented optical lattices~\cite{Anderson1998}.
This gauge correspondence permits the Bloch dynamics to be described in terms of the static tilted lattice, where the intraband motion is most transparently formulated in momentum space and is subsequently mapped onto a discrete tight-binding lattice representation.

\subsection{Characteristic and Dimensionless Scales}
To facilitate the analysis, we introduce a set of characteristic scales associated with the optical lattice and express the governing equation in dimensionless form.
The characteristic energy scale is chosen as $\tilde{E} = 4\hbar^2 k_L^2/M$,  which is related to the recoil energy $E_r = \hbar^2 k_L^2/2M$, with $k_L = 2\pi/\lambda_L$ denoting the wavenumber of the laser field with wavelength $\lambda_L$.
Correspondingly, the characteristic length, time, force, and wavefunction scales are given by $\tilde{x} = 1/2k_L$, $\tilde{t} = \hbar/\tilde{E} = M/4\hbar k_L^2$, $\tilde{F} = 8\hbar^2 k_L^3/M$, and $\tilde{\Psi} = \sqrt{n_0}$, where $n_0$ is the average density of the condensate and has the unit of inverse length.

We start in a quasi-one-dimensional regime realized by strong transverse harmonic confinement with $\omega_y=\omega_z\equiv\omega_\perp$, such that the transverse degrees of freedom are frozen and can be integrated out. The resulting strength of effective one-dimensional two-body interaction is
$U_{1}=U^{(0)}/(2\pi a_\perp^2)$, where $U^{(0)}=4\pi\hbar^2 a_s/M$ and
$a_\perp=\sqrt{\hbar/(M\omega_\perp)}$. For symmetric binary mixtures forming liquid droplets, the mean-field nonlinearity is governed by the residual interaction imbalance between intra- and interspecies couplings, while the LHY correction originates from the quantum fluctuations~\cite{Petrov2016}. Consequently, the two effective nonlinearities scale as
$\propto \hbar^2 a_s/(M a_\perp^2)$ and $\propto \hbar^2 a_s^{3/2}/(M a_\perp^3)$, respectively. Expressed in the dimensionless scales introduced
above, the mean-field and LHY nonlinearities are encoded in two coefficients $c_L$ and  $d_L$, whose reference values are $c_L^{(0)}=a_s n_0/(2a_\perp^2 k_L^2)$ and $d_L^{(0)}=\sqrt{n_0}\,a_s^{3/2}/(\pi k_L^2 a_\perp^3)$.

\subsection{GPE in Tight-Binding Model}
Rescaling the spatial, temporal, and wavefunction variables according to the characteristic units defined above, the time-dependent Gross-Pitaevskii equation (GPE) can be written in the following dimensionless form:
\begin{align}
i\frac{\partial\Psi(x,t)}{\partial t} &= -\frac{1}{2}\frac{\partial^2 \Psi(x,t)}{\partial x^2} + V_{\rm ext}(x)\Psi(x,t) + c_L |\Psi(x,t)|^2\Psi(x,t) - d_L |\Psi(x,t)|\Psi(x,t).
\label{Eq-1}
\end{align}
As a counterbalance of the mean-field interaction, the presence of the LHY correction and the competition between them greatly reshape the stationary and dynamic properties of the quantum droplet in a nontrivial manner. In the absence of the nonlinear terms and external driving, we consider stationary solutions $\Psi(x,t) = e^{-i E t}\Phi(x)$, which leads to the linear Schr\"{o}dinger eigenvalue problem for a 1D periodic potential with a constant strength $V_0$,
\begin{align}
E \Phi(x) &= -\frac{1}{2}\frac{d^2 \Phi(x)}{d x^2} + V_0\sin^2\left({\pi x}/{2}\right)\Phi(x). 
\label{Eq-2}
\end{align}
%
The eigenfunctions of Eq.~(\ref{Eq-2}) can be expressed in terms of the Mathieu functions. Although originally introduced in the context of classical vibration problems such as elliptical membranes, Mathieu equations arise naturally when the 1D Schrödinger equation is written for a sinusoidal periodic potential, which have been extensively employed in a variety of related contexts, including quadrupole ion traps \cite{2March1997,4Baranov2003}, ultracold atomic lattices \cite{5Rey2005}, and quantum rotor models \cite{Ayub2012}.

Under Born-von Karman periodic boundary conditions, the eigenstates take the Bloch form $\Phi_k(x) = e^{ikx}u_k(x)$, where $u_k(x)$ shares the lattice periodicity, i.e., $u_k(x+2)=u_k(x)$. Expanding the periodic part as a Fourier series
and substituting it into Eq.~(\ref{Eq-2}) yields a tridiagonal Bloch-band eigenvalue problem. 
In the deep-lattice regime, the Bloch bands become narrow for energies well below the potential maxima, reflecting the increase in localization within the individual sites. While Bragg scattering persists for any nonvanishing periodic potential, the low-energy dynamics are dominated by the lowest band. Near each potential minimum, the sinusoidal potential can be locally approximated by a harmonic oscillator. The resulting eigenfunctions provide a natural basis for describing the localized states, motivating the use of Wannier functions for the single-band dynamics.

Assuming a weak $k$-dependence of $u_k(x)$ and neglecting the site-specific phase shift, we approximate the Wannier function at each site $n$ as $W_n(x) \simeq \mathrm{sinc}\!\left(\pi x/2\right)\,u(x)$.
Enforcing the orthonormality within single-band approximation allows the decomposition $\Psi(x,t) = \sum_n \psi_n(t) W_n(x)$ to yield an effective discrete GPE, 
\begin{align}
i \frac{\partial \psi_n(t)}{\partial t}
&= \varepsilon_n \psi_n(t)
- J \bigl[\psi_{n+1}(t) + \psi_{n-1}(t)\bigr]
+ A |\psi_n(t)|^2 \psi_n(t)
- B |\psi_n(t)| \psi_n(t),
\label{Eq-discGPE}
\end{align}
where the on-site energy $\varepsilon_n
= \langle W_n | H_{\rm lin} | W_n \rangle$ $\equiv$ $\varepsilon_n^{(0)} + Fn$
and the nearest-neighbor hopping amplitude $J = \langle W_n | 
H_{\rm lin} | W_{n+1} \rangle$
are determined by the linear part of the Hamiltonian $H_{\rm lin}$. 
The nonlinear coefficients are given by the overlap integrals $A = \int|W(x)|^4\, dx $ and $B = \int |W_n(x)|^3\,dx $.  
Via the Legendre transformation, we identify the canonical variables and derive the total Hamiltonian as:
\begin{align}
\mathcal{H}(t) = \sum_n \left[
-J(\psi_n\psi_{n+1}^* + c.c.)
+ \varepsilon_n|\psi_n|^2
+ \frac{A}{2}|\psi_n|^4
- \frac{2B}{3}|\psi_n|^3
\right].
\label{Eq-Htot}
\end{align}
This formulation provides the basis for analyzing the intraband dynamics and localization phenomena of quantum droplets in optical lattices.

Throughout this work, we choose $c_L=1$ as a representative normalization of the mean-field nonlinearity, and consider $d_L=0$ and $1$ to describe respectively the switch-off and switch-on of the LHY contribution while keeping the mean-field nonlinearity unchanged. We note that the mean-field and LHY terms remain distinct in their functional dependence and therefore do not cancel even when their prefactors are comparable. For a deep lattice, $V_0=10$, we obtain $A=3.412$, $B=2.44$, $J=0.457$, and $\varepsilon_0=6.964$ in the absence of an external force, i.e., $F = 0$.

\section{Breathing, Self-Trapping, and Bloch Oscillation} 

The stationary properties and the dynamical response of the droplets to external forces are investigated using a time-dependent variational approach. We construct an effective Lagrangian from the discrete Hamiltonian and derive the equations of motion for the variational parameters $\{R, \alpha, p, \delta\}$ using a generalized super-Gaussian ansatz:
\begin{align}
\psi_n(t)
= \sqrt{\rho\,N}\,\exp\big[{- z_n^{2m}\,/\,\alpha^{2m}}\big] \exp\big[\,{ i p z_n }\,\big] \exp \big[\,{ i \delta z_n^2 \,/\, 2 }\,\big],  
\end{align}
where $z_n = n - R$ and $\rho = 2^{-2m}\,\alpha^{-1}\,m\,\Gamma\!\left({m^{-1}}/{2}\right)$ is the normalization constant ensuring the conservation of the scaled particle number $N$. Here \(R(t)\) and \(\alpha(t)\) denote the center-of-mass position and wavepacket width, while \(p(t)\) and \(\delta(t)\) denote the quasi-momentum and the quadratic phase curvature of the chirp, respectively.
The exponent \(m\) is a continuous shape parameter characterizing the generalized super-Gaussian envelope.
The internal phase structure encoded in the chirp is essential to the wavepacket deformation in the system with nonlinear and dispersive effects, and corresponds to the local quasimomentum defined through the discrete phase gradient,
\begin{align}
k_{\rm loc}(n)\simeq \frac{\phi_{n+1}-\phi_{n-1}}{2} = p + \delta z_n. \label{Eq-kloc}
\end{align}
Thus, a positive chirp means that the right and left sides of the droplet carry larger and smaller local quasimomenta, respectively, while a negative chirp reverses this position-momentum correlation. Depending on the sign of the lattice dispersion, determined by the local band curvature, the same chirp may correspond to a tendency toward compression, expansion, or dephasing-induced suppression of mobility.

This ansatz extends the standard Gaussian wavepacket $(m = 1)$ description by allowing for a tunable envelope shape that interpolates among sharp sub-Gaussian, Gaussian, and flat-top super-Gaussian profiles.
The Gaussian-based variational approaches, including linear and quadratic phase terms, have been widely used to describe collective transport, self-trapping, and nonlinear Bloch dynamics in lattice Bose-Einstein condensates and discrete nonlinear Schr\"odinger models \cite{Trombettoni2001, Anderson1998, KevrekidisBook}. On the other hand, the super-Gaussian envelope describes slowly varying amplitudes \(\psi_n(t)\), whereas the sub-Gaussian describes sharp and localized amplitudes, while lattice-scale spatial modulation is contained in the Wannier functions. This ansatz is therefore appropriate as long as the envelope remains smooth, single-peaked, and extends over several lattice sites $(\alpha \ge 2)$; it may lose quantitative accuracy in regimes dominated by strong discreteness or multi-peak fragmentation {\cite{Hennig1998, KevrekidisBook}}.

Substituting the ansatz into the Lagrangian and performing the lattice summations yields the effective Lagrangian per particle:
\begin{align}
\mathfrak{L}
= p \mkern1.5mu \dot{R}
- a(m) \mkern1.5mu \alpha^2 \mkern1.5mu \dot{\delta}
+ b(m) \mkern1.5mu \alpha^{m-1} \,\mkern1.5mu e^{-\eta} \mkern1.5mu \cos p
- c(m) \mkern1.5mu \alpha^{-1} \mkern1.5mu N
+ d(m) \mkern1.5mu \alpha^{-1/2} \mkern1.5mu N^{1/2}
- \varepsilon_0
- R \mkern1.5mu F ,  \label{Eq:Lagranian}
\end{align}
where 
\begin{align}
\eta(\alpha,\delta) = {2^{1-2m}}{\alpha^{-2m}} + 2^{2m-4}(2m)^{-1}(2m-1)^{-1}{\delta^2 \alpha^{2m}}  \label{Eq-eta}
\end{align}
and the coefficients $a(m)$, $b(m)$, $c(m)$, and $d(m)$ are defined in Appendix~\ref{App:deep-abcd}. Equation~(\ref{Eq:Lagranian}) identifies $(R, p)$ and $(a(m)\alpha^2, \delta)$ as pairs of canonically conjugate variables. Additionally, the term proportional to $\cos p$ describes the phase-modulated, Bloch-driven inter-cell hopping, and the terms with particle number weightings capture the envelope-squeezing enhancement in the mean-field and LHY nonlinear effects.  
The resulting Euler-Lagrange equations yield the evolution of the collective variables:
\begin{eqnarray}
\dot{R} &=& b(m) \mkern1.5mu \alpha^{m-1} \mkern1.5mu e^{-\eta} \mkern1.5mu \sin p, \label{Eq-xi} \\
\dot{p} &=& -F, \label{Eq-p} \\
\dot{\alpha}
&=& \frac{b(m)}{a(m)}\,
\frac{2^{2m-5}\,\alpha^{3m-2}}{m\,(2m-1)}\,\delta\, e^{-\eta}\,\cos p, \label{Eq-alpha} \\
\dot{\delta}
&=& \frac{c(m)}{2 a(m)}\,\alpha^{-3}\,N
- \frac{d(m)}{4 a(m)}\,\alpha^{-5/2}\,N^{1/2}
+ \left[ \frac{b(m)}{2 a(m)}\,\alpha^{m-3}\, e^{-\eta}\,\left(m-1-\alpha\, \frac{\partial \eta}{\partial \alpha}\right)\right]\,\cos p. \label{Eq-delta}
\end{eqnarray}
While Eq.~(\ref{Eq-p}) explicitly shows the linear response of quasimomentum to a constant external force, the relation $p(t)=p_0-F t$ enters the remaining equations of motion sinusoidally. This encoding implies that the system dynamics depend on both the elapsed  time and the initial value $p_0$. Therefore, the specific points at $p_0=0$ and $p_0=\pi / 2$ offer ideal baselines for examining the internal competitions in Bloch dynamics with and without the external force. Further, a direct proportion between $\dot{\alpha}$ and $\delta$ in Eq.~(\ref{Eq-alpha}) validates the dispersion compensation from chirp on the wavepacket expansion or compression. 
Moreover, Eq.~(\ref{Eq-delta}) indicates that the chirp rate is crucial to the Bloch dynamics by showing that it can be separated into a nonlinear contribution and a Bloch-modulated lattice-dispersion contribution, $\dot{\delta} \simeq A_{\rm NL}(\alpha) + B_{\rm lat}(\alpha,\delta)\cos(p_0-F t)$. Here $A_{\rm NL}$ comes from the nonlinear balance between mean-field and LHY interactions that is independent of the quasimomentum, whereas the second term originates from the hopping-induced lattice dispersion $B_{\rm lat}$ and becomes periodically modulated by the Bloch sweep of the quasimomentum.

After some algebra, the center-of-mass motion is further found to obey an effective driven oscillator equation
\begin{align}
\ddot{R} + \gamma \dot{R} + F^2 R
= \mathcal{F}_{\mathrm{eff}}, \label{Eq-dampEOM}
\end{align}
governed by the time-dependent effective driving force $\mathcal{F}_{\mathrm{eff}} = F \mathcal{H}_{\mathrm{eff}}(0) - N F \alpha^{-1} c(m)
+ \sqrt{N} F \alpha^{-1/2} d(m)$, and the nonlinear damping-like effect through the parameter
\begin{align}
\gamma(t) = \delta(t)\mathcal{G}[\alpha(t)],  \label{Eq-gamma}
\end{align}
where $\mathcal{G}[\alpha] \propto  A_{\rm NL}(\alpha)$. While this term explicitly arises from the multiplicative combination of internal phase deformation and nonlinear competition, it does not always imply positive resistance. Rather, an anti-damping mechanism could be triggered by the intrinsic feedback via an implicit nonlinear gain. 

In accordance with Hamilton's principle of least action, an effective Hamiltonian $\mathscr{H}_{\mathrm{eff}}$ can be established using Legendre transformation. The group velocity and the inverse effective mass are then determined as follows:
\begin{align}
V_g &= \partial\mathscr{H}_{\mathrm{eff}}/{\partial p} = b(m)\, \alpha^{m-1} \,e^{-\eta} \, \sin p \nonumber\\ 
1/M^{*} &= \partial^2 \mathscr{H}_{\mathrm{eff}}/{\partial p^2} = b(m)\, \alpha^{m-1} \, e^{-\eta} \, \cos p,  \label{Eq-VgMstar}
\end{align}
consistent with the dynamical relation $\dot{R} = \tan p / M^*$. 
Together with Eq.~(\ref{Eq-kloc}), the local velocity of the wavepacket at a lattice displacement $z$ from its center can be approximated by expanding the band velocity around the central quasimomentum $p$:
\begin{align}
v(k_{\rm loc})  \approx v(p) + \mathscr{H}^{''}_{\mathrm{eff}}(p) (k_{\rm loc} - p)  = v(p) + \delta\,z\,/M^*. 
\end{align} 
This expression shows explicitly that for $z > 0$, the local velocity exceeds the center-of-mass velocity, $v(k_{\rm loc}) > v(p)$, when $\delta\,z\,/M^* > 0$, whereas it becomes smaller than the center-of-mass velocity, $v(k_{\rm loc}) < v(p)$, when $\delta\,z\,/M^* < 0$. The same criterion applies with corresponding sign reversal on the opposite side of the wavepacket. Therefore, the tendency toward wavepacket broadening or narrowing is explicitly governed by  
\begin{align}
\mathscr{W}_{\pm} = \operatorname{sgn} \left(\delta \cos p\right), \label{Eq-sgn}
\end{align}
which represents the combined effect of the internal phase curvature and the local band dispersion.
This observation reveals that the mean-field and LHY nonlinearities, together with their mutual competition, are nontrivial and delicate in determining the wavepacket deformation and dynamics through the directional gradient driving of the internal phase degrees of freedom in Eqs.~(\ref{Eq-xi})--(\ref{Eq-delta}).

At $t=0$, we set the initial conditions $p=p_0$, $R=0$, $\delta=0$, and $\alpha=\alpha_0$. 
The initial width $\alpha_0$ and the generalized super-Gaussian parameter $m$ can be determined by minimizing the stationary effective Hamiltonian $\mathscr{H}_{\mathrm{eff}}(0)$, thereby defining the droplet profile prior to the kick. For a deep lattice with $V_0=10$ and $N = 1$, this minimization gives us $\alpha=3.194$ and $m=0.910$ for the case $c_L=1$ and $d_L=1$, and $\alpha=7.145$ and $m=0.789$ for the case $c_L=1$ and $d_L =0$. We find that the profiles for both cases not only possess sub-Gaussian characteristics $(m < 1)$ but also remain single-peaked extending over several lattice sites $(\alpha > 2)$. To be more precise, these two envelopes well capture the characteristics of optical solitons described by $\sech(0.5 t)$ and $\sech(0.23 t)$, respectively, and therefore are suitable for investigating the transport phenomena in the tight-binding model.   


%
The collective dynamics of the quantum droplets arises from the interplay between lattice dispersion and nonlinear interactions. The inverse effective mass controls the sign and strength of dispersion, while the net nonlinearity determines the self-confinement tendency and internal phase evolution. Depending on the initial conditions and driving protocols, this competition, together with a phase-induced nonlinear damping-like parameter, gives rise to several characteristic dynamical responses: kinematic freezing of the center of mass, internal breather-like oscillations, chirp-induced arrest of finite-momentum transport, and coherent damping of nonlinear BO. Summarized in Fig.~\ref{Fig1}, with the blue curves obtained by systematically switching off the LHY correction and the red curves showing the trajectories with full mean-field and LHY effects, these responses provide the reference framework for more detailed dynamics examined in the following subsections.







\subsection{Kinematic Freezing and LHY-Activated Internal Breathing at $p_0 = 0$}

\begin{figure*}[t!]
\includegraphics[width=0.94\textwidth]{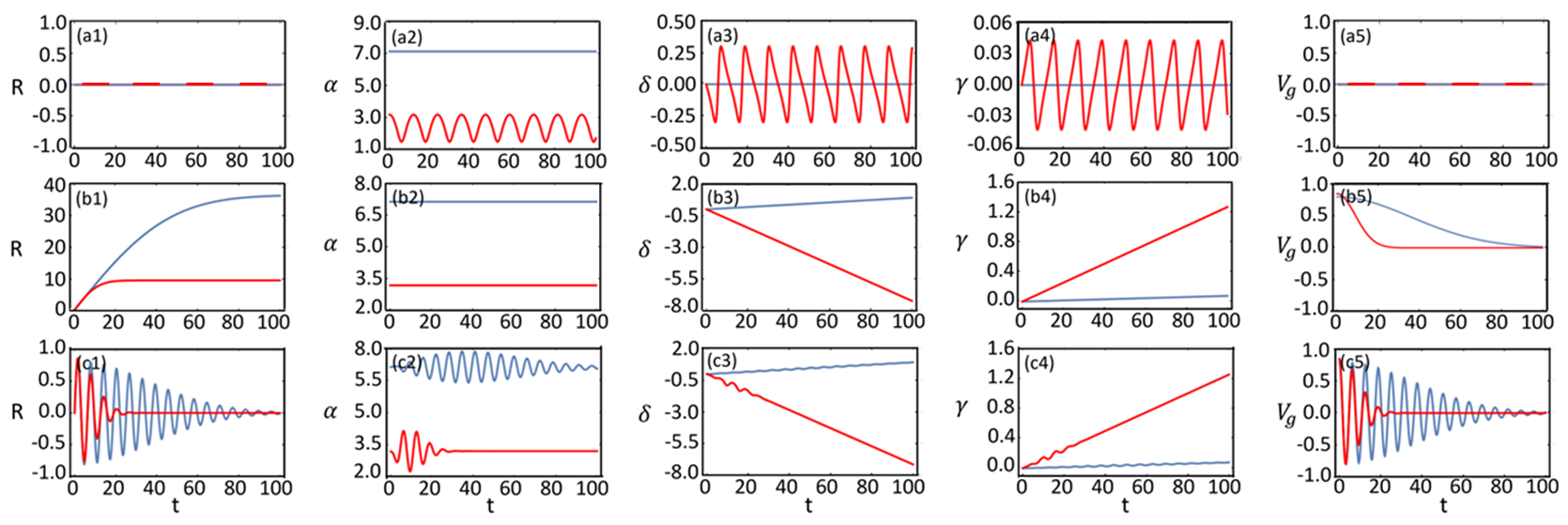}
\caption{(color online) Evolution of the wavepacket collective variables for a scaled particle number $N = 1$ in three representative regimes: (a) $F=0$ and $p_0 =0$; (b) $F =0$ and $p_0 = \pi/2$; and (c) $F=1$ and $p_0 =\pi/2$. Red curves represent the full dynamics including both the mean-field and LHY nonlinearities, while the blue curves denote dynamics where the LHY term is systematically deactivated. The columns illustrate the competition between lattice dispersion, nonlinearities, internal phase deformation, and external forces, which collectively govern the transition among breather oscillations, self-trapping, and nonlinear Bloch damping.
}\label{Fig1}
\end{figure*}

As shown in Fig.~\ref{Fig1}(a), when the external force is absent $(F = 0)$, the quasimomentum is conserved with $p(t) = p_0$ throughout the evolution. At $p=0$, corresponding to the $\Gamma$ point at a band extremum, Eq.~(\ref{Eq-VgMstar}) gives a vanishing group velocity when $\sin p = 0$. The center-of-mass motion is therefore frozen at $R=0$. As a consequence, both the position $R$ and group velocity $V_g$ remain zero in time, as evidenced by the complete overlap of the red and blue curves in Fig.~\ref{Fig1}(a1) and (a5). This behavior is a direct consequence of quasimomentum conservation rather than a result of nonlinear self-trapping.

Although the translational motion is frozen by the conserved quasimomentum, the internal degrees of freedom remain sensitive to nonlinear interactions. In the absence of the LHY correction (blue curves), the width parameter $\alpha$ remains constant, indicating that the wavepacket maintains a stationary profile. In contrast, when the LHY term is retained (red curves), the center-of-mass motion remains frozen, but $\alpha$ exhibits persistent oscillations accompanied by corresponding oscillatory behavior in $\delta$ and $\gamma$ [Fig.~\ref{Fig1}(a2)-(a4)]. This shows that the LHY correction does not generate translational localization in this case, but activates an internal breathing response within a kinematically frozen background.

The phase relation among the internal variables further clarifies the nature of this breathing mode. For $p=0$, Eq.~(\ref{Eq-alpha}) reduces schematically to $\dot{\alpha}\propto \delta$, so that the chirp plays the role of a velocity-like variable conjugate to the width. The driving mechanism is provided by Eq.~(\ref{Eq-delta}), the sum of $A_{\rm NL}(\alpha)$ and $B_{\rm lat}(\alpha,\delta)$. Therefore, a finite chirp can be triggered at $t>0$, requiring a nonzero $\dot{\delta}(0)$ at $t = 0$ even though $\delta(0) = 0$.

Given the specific parameters $\alpha$ and $m$ for the initial envelope profiles, we find that $B_{\rm lat} < 0$, where $A_{\rm NL}(\alpha) + B_{\rm lat} \sim -0.06$ with LHY, but approaches to $10^{-15}$ without LHY. The balance between wavepacket compression and expansion, fully compensated by the negative lattice dispersion and the mean-field nonlinearity, halts the dynamics of the blue trajectories.
However, the inclusion of the LHY term reduces the inherent mean-field expansion, allowing for an imbalance between the lattice dispersion and the nonlinearity. To be more precise, $\dot{\delta}(t)$ develops a spiky pulse train with secondary crests inside the troughs. This acceleration-like response excites the chirp and subsequently drives the width oscillation. Therefore, a bounded exchange between $\alpha$ and $\delta$ produces an internal breathing oscillation with a finite phase lag between the density width and the quadratic phase curvature.

While a true soliton-like balance would require a stationary relation between dispersion and nonlinearity, together with a phase structure that does not accumulate secularly, the wavepackets represented by either the red or the blue curves are therefore not propagating solitons, but initially mobile localized states whose phase curvature progressively modifies their transport.

\subsection{Finite-Momentum Transport and Chirp-Induced Mobility Arrest at $p_0=\pi/2$}

When the conserved quasimomentum is chosen near the Brillouin zone edge at $p = p_0 = \pi/2$, the situation changes qualitatively as shown in  Fig.~\ref{Fig1}(b). Based on Eq.~(\ref{Eq-VgMstar}) and $\sin p=1$, the initial group velocity is nonzero, and the droplet therefore undergoes center-of-mass transport even in the absence of an external force. At the same time, $\cos p=0$ makes the inverse effective mass vanish for a flat band, suppressing the usual lattice-assisted compression. Consequently, given $\dot{\alpha} = 0$, the wavepacket width remains constant over time as expected from the expression $\mathscr{W}_{\pm} = \operatorname{sgn} \left(\delta \cos p\right) = 0$ in Eqs.~(\ref{Eq-alpha}) and (\ref{Eq-sgn}), and verified in Fig.~\ref{Fig1}(b2).

The remaining long-time evolution is governed by the phase-curvature dynamics, which renormalizes the effective mobility through the factor $\eta(\alpha,\delta)$. As shown in Eq.~(\ref{Eq-delta}), it is clear that the chirp dynamics is now governed by $A_{\rm NL}(\alpha)$, which is a constant in time without the wavepacket deformation. 
This leads to a linear chirp accumulation following $\delta(t) \sim \lambda t$, as shown in Fig.~\ref{Fig1}(b3). Accordingly, a gradually growing positive chirp induced fully by the mean-field effect is built on the blue trajectory, whereas LHY correction turns the effective nonlinear interaction to an attractive effect, which changes the position-momentum correlation inside the droplet and builds a rapidly-growing negative chirp on the red trajectory. These two trajectories do not simply differ in magnitude; they correspond to different internal phase-building routes determined by the initial acceleration $\dot{\delta}(0)$.

Regardless of the sign of the accumulated linear chirp, the mobility-arrest effect is triggered when $\eta(t) \sim \lambda^2 t^2$ enters $e^{-\eta}$ as the accumulated quadrature chirp. Consequently, the droplet reduces its mobility in a Gaussian-like decay and completely stops in the long-term limit, as in Fig.~\ref{Fig1}(b5). In parallel, the chirp-induced translation arrest can be analytically observed in Eq.~(\ref{Eq-xi}). Since $\dot{R} = V_g$, it is simple to obtain $R(t)$ as an error function ${{\rm erf}(t)}$ with a saturation profile at large $t$, a feature precisely displayed in Fig.~\ref{Fig1}(b1). Overall, the LHY correction here does not replace self-trapped transport by diffusion; instead, it redirects the internal phase route through which the arrest of an initially mobile droplet is accelerated.


\subsection{Nonlinear Phase Modulation, Chirp Reweighting, and the Optical-SPM Analogy}

The two force-free limits discussed above show that the nonlinear interaction does not act primarily as a direct force on the wavepacket center. Instead, it reshapes the internal phase structure through the chirp variable \(\delta\). This role is analogous to the self-phase modulation (SPM) in nonlinear optics, where an intensity-dependent refractive index produces a position-dependent nonlinear phase across an optical pulse. For an optical envelope \(A_{\rm opt}(T,z)=\sqrt{P(T,z)}e^{i\phi(T,z)}\), where $T$ is the retarded time in the co-moving frame, and $P$ is the optical power, the Kerr nonlinearity generates a nonlinear phase \(\phi_{\rm NL}\propto \gamma_{\rm K}P(T)z\). Here, \(\gamma_{\rm K}\) denotes the Kerr nonlinear coefficient, proportional to the nonlinear refractive index \(n_2\). Since the instantaneous frequency shift is obtained from the phase gradient, \(\Delta\omega = -\partial_T\phi_{\rm NL}\), the nonlinear phase modulation converts the intensity profile into a chirp. The resulting pulse broadening or compression is then determined not by the chirp alone, but by the product of the chirp slope and the group-velocity dispersion.

A directly corresponding mechanism appears in the present lattice droplet. The quadratic phase in the envelope ansatz gives the local quasimomentum \(k_{\rm loc} = p + \delta z\), as shown in Eq.~(\ref{Eq-kloc}), so that \(\delta\) measures the spatial chirp of the droplet rather than an external force. The subsequent deformation is selected by the local band curvature, since the tendency toward broadening or narrowing is governed by Eq.~(\ref{Eq-sgn}), \(\mathscr{W}_{\pm} = \operatorname{sgn}(\delta\cos p)\). In this sense, the pair \(\delta\) and \(1/M^*\propto\cos p\) plays the same structural role as the chirp slope and group-velocity dispersion in chirped-pulse propagation. The sign of \(\delta\) determines the position-quasimomentum correlation, while the sign of \(1/M^*\) determines how this correlation is converted into expansion or compression. 

For an intuitive analogy to the SPM effect, the mean-field nonlinearity in the GPE provides the familiar nonlinear phase-modulation channel, through which an internal phase-gradient drive can be encoded in the chirp variable \(\delta\), while the resulting deformation is selected by the local band curvature through \(\mathscr{W}_{\pm}\). However, the LHY correction makes this phase-modulation mechanism nontrivial. It is not simply a static counterterm against the mean-field interaction. Owing to its different density dependence and opposite sign in the effective nonlinear phase rate, the LHY contribution penetrates the chirp-generation channel itself. It can therefore reweight, suppress, or reverse the mean-field induced phase-gradient drive. 

This behavior is reminiscent of nonlinear pulse-shaping mechanisms in cubic-quintic Ginzburg-Landau systems \cite{kuan2018}, where higher-order or nonstandard nonlinear responses may become decisive by modifying the effective shaping channel rather than by simply adding a small perturbative correction. In such systems, pulse states can be selected by the balance among dispersion, SPM, and higher-order saturable absorption. Analogously, in the present conservative droplet dynamics, the LHY term acts as a nonstandard nonlinear phase response. It does not merely shift the static energy balance. Instead, it percolates into the internal phase dynamics, reshapes the mean-field induced phase-gradient drive and may become decisive in regimes where the mean-field and lattice-dispersion contributions nearly compensate each other.

The damping-like coefficient \(\gamma(t)=\delta(t)\mathcal{G}[\alpha(t)]\) further clarifies how the internally generated chirp is converted into an effective feedback on the collective motion. Since \(\mathcal{G}[\alpha]\) is determined by the nonlinear balance between the mean-field and the LHY interactions, \(\gamma\) is not a direct copy of the chirp. Rather, it represents the chirp after being reweighted by the nonlinear response function. In this sense, the interpretation of \(\gamma\) is closer to nonlinear pulse-shaping theory than to pure Kerr SPM alone. In the Kerr limit, the nonlinearity first generates a chirp through an intensity-dependent phase. In the cubic-quintic Ginzburg-Landau descriptions of mode-locked lasers, however, the consequence of this phase modulation is further selected by nonlinear shaping channels, such as cubic and quintic saturable absorption. The analogy here is not that the LHY term acts as a dissipative saturable absorber, but that a nonstandard nonlinear response can reshape the effective dynamical channel through which phase modulation acts.

This distinction is particularly important for the force-free case \(p_0=0\). In this limit, the center-of-mass motion is kinematically frozen because \(\sin p=0\), so the term \(\gamma\dot{R}\) in Eq.~(\ref{Eq-dampEOM}) is not damping an active translational motion. The oscillatory behavior of \(\gamma\) in Fig.~\ref{Fig1}(a4), including its negative portions, should therefore not be interpreted as an ordinary anti-damping instability of \(R(t)\). Instead, it marks the compression-feeding part of the internal chirp-width cycle. The LHY-induced imbalance drives the chirp variable \(\delta\), Eq.~(\ref{Eq-xi}) converts this chirp into the breathing motion of the width through \(\dot{\alpha}\propto\delta\cos p\), and the nonlinear response function \(\mathcal{G}[\alpha]\) determines whether the resulting phase curvature enters the collective feedback with a damping-like or antidamping-like sign. Thus, a negative \(\gamma\) indicates a coherent return of internal phase deformation into the width dynamics, rather than a thermodynamic gain mechanism.

For the finite-momentum case \(p_0=\pi/2\), the same reweighting mechanism becomes visible in the transport dynamics because the wavepacket carries a nonzero group velocity. In this case, \(\gamma\) diagnoses how the chirp acts back on the center-of-mass motion after being filtered by the mean-field--LHY response. A negative chirp can still produce a positive damping-like feedback when \(\mathcal{G}[\alpha]<0\), as occurs in the LHY-modified trajectory shown in Fig.~\ref{Fig1}(b4). Therefore, the sign of \(\gamma\) should not be identified with the sign of \(\delta\) alone. It measures how the internal phase curvature is reweighted before it suppresses or reinforces the coherent motion of the droplet.

\subsection{Nonlinear BO under External Force}

%
%


As shown in Fig.~\ref{Fig1}(c), the application of a constant external force \(F=1\) drives the wavepacket into Bloch-type dynamics. In contrast to the force-free cases, where the quasimomentum is conserved, the external force continuously sweeps \(p(t)=p_0-Ft\) across regions of different band curvature. The stability and eventual suppression of BO are therefore controlled not only by the center-of-mass motion, but also by the coupled evolution of the width \(\alpha\) and the chirp \(\delta\).

For the trajectory without the LHY correction, the initial coefficients entering the chirp equation are
\(A_{\rm NL}\simeq 9.228228\times10^{-3}\) and
\(B_{\rm lat}\simeq -9.228228\times10^{-3}\). 
Since \(p_0=\pi/2\) and \(F=1\), the Bloch-modulated factor becomes
\(\cos(p_0-Ft)=\sin t\), and the early-time chirp dynamics can be estimated as $\dot{\delta}(t)\simeq 0.009228228-0.009228228\sin t$. Together with the initial condition \(\dot{\delta}(0)\simeq 0.009228228>0\), this shows that the chirp is first driven by a positive nonlinear bias. As the external force sweeps the quasimomentum away from \(p_0=\pi/2\), the lattice-dispersion contribution grows with the opposite sign and nearly compensates the nonlinear drive. This near cancellation prevents a rapid secular accumulation of chirp, so that the blue trajectory in Fig.~\ref{Fig1}(c3) remains dominated by small oscillatory ripples rather than by a strong drift.

For the LHY-modified trajectory, the corresponding coefficients become
\(A_{\rm NL}\simeq -1.1313692\times10^{-2}\) and
\(B_{\rm lat}\simeq 1.1313459\times10^{-2}\). 
The short-time chirp equation is therefore estimated as $\dot{\delta}(t)\simeq -0.011313692+0.011313459\sin t$, with \(\dot{\delta}(0)\simeq -0.011313692<0\). 
Thus, the LHY correction reverses the initial direction of chirp generation relative to the trajectory without LHY. In the language of the preceding subsection, the LHY term does not merely weaken the mean-field response; it changes the sign of the nonlinear phase-gradient drive and redirects the position-quasimomentum correlation inside the droplet. Although the nonlinear bias and the Bloch-modulated lattice-dispersion term again appear with nearly opposite signs at early times, this compensation is not sustained during the subsequent evolution. As $\alpha$, $\eta$, and $\delta$ become time dependent, the LHY-modified nonlinear response is continuously reweighted, leaving a finite period-averaged chirp drive. This residual drive produces the long-time secular drift observed in the red trajectory of Fig.~\ref{Fig1}(c3).

The nonlinear filtering discussed above is quantified by the coefficient $\mathcal{G}[\alpha]$ in Eq.~(\ref{Eq-gamma}).
Using the initial parameters, we obtain $\mathcal{G}\simeq 0.08405419$ for the trajectory without the LHY correction, but \(\mathcal{G}\simeq -0.02769830\) when the LHY correction is included. 
Thus, for the blue trajectory, the early-time feedback follows $\gamma(t)\simeq 0.08405419\,\delta(t)$,
so that \(\gamma\) carries the same sign as the chirp. 
For the red trajectory, however, the balance between the mean-field and LHY interactions reverses the response coefficient and gives
$\gamma(t)\simeq -0.02769830\,\delta(t)$.
Since the LHY-modified chirp is initially driven toward negative values, this negative response coefficient converts the negative chirp into a positive damping-like feedback. 
Therefore, the positive \(\gamma\) observed in Fig.~\ref{Fig1}(c4) is not in contradiction with the negative $\delta$ in Fig.~\ref{Fig1}(c3). 
It is the direct quantitative signature of how the mean-field and LHY responses convert the chirp into an effective feedback.
Together with the slower width evolution in Fig.~\ref{Fig1}(c2), this positive damping-like feedback reduces the amplitude of the center-of-mass BO through the effective term \(\gamma\dot{R}\) in Eq.~(\ref{Eq-dampEOM}). 
The suppression is therefore not set by the sign of the chirp alone, but by the sign of the effective product $\delta\,\mathcal{G}[\alpha]$.

The suppression of BO also appears through the mobility-renormalization factor in Fig.~\ref{Fig1}(c1) and (c5). 
While \(\gamma(t)=\delta(t)\mathcal{G}[\alpha(t)]\) retains the sign-sensitive feedback of the chirp, the factor \(e^{-\eta}\) provides a sign-insensitive suppression channel. 
From Eq.~(\ref{Eq-eta}), the chirp-dependent part of \(\eta\) can be written as $\eta=\eta_0 + C_{\eta}\delta^2$, with \(C_{\eta}=4.554189291\) and \(1.224105599\) for the trajectories without and with LHY correction, respectively. 
Thus, although the two trajectories develop chirps with different signs and are filtered by different nonlinear response coefficients, the reduction of mobility is controlled by the accumulated magnitude $|\delta|$, rather than by the sign of $\delta$. As $|\delta|$ grows, the factor $e^{-\eta}$ progressively reduces the effective group velocity and weakens the oscillatory response. 
The center-of-mass motion therefore evolves from underdamped BO toward a progressively arrested trajectory. 
This damping is not dissipative in the thermodynamic sense; it reflects the coherent transfer of center-of-mass motion into internal phase deformation.

\begin{figure}[h!]
\includegraphics[width=0.45\textwidth]{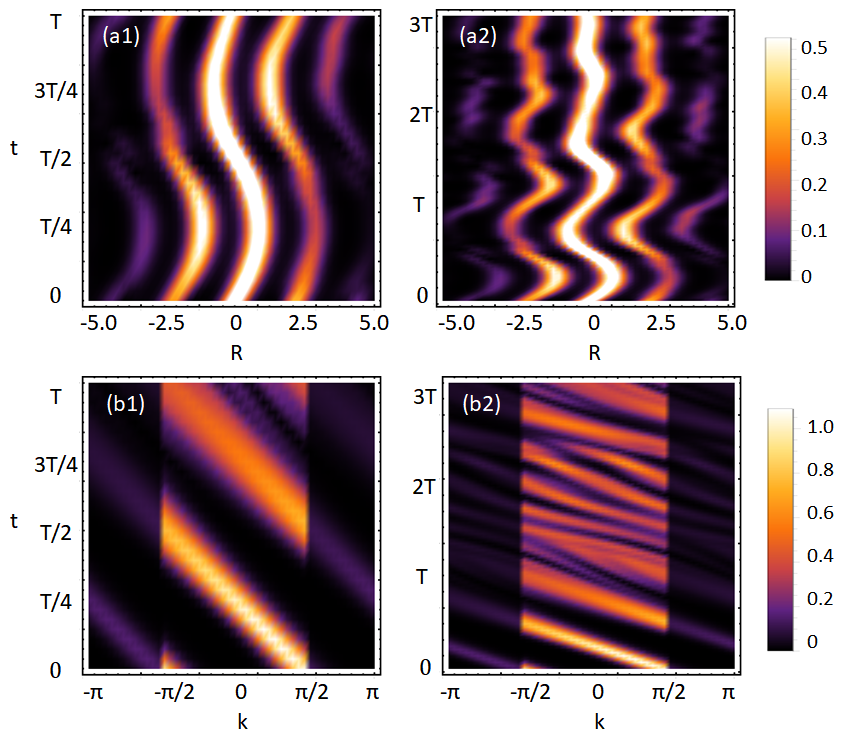}
\caption{(color online) Time evolution of the condensate density in real space $|\Psi(x, t)|^2$[(a1),(a2)] and momentum space $|\Psi(k, t)|^2$[(b1),(b2)] in the presence of LHY correction, for $F=1$ and $p(0)=\pi / 2$. Panels (a1) and (b1) demonstrate the dynamics within the first Bloch period $T \sim 6.29$, while (a2) and (b2) show the long-time evolution up to $3 T$. The coherent Bloch oscillations observed for $t < T$ gradually degrade due to nonlinear dephasing, leading to spatial broadening and momentum-space fragmentation.
}\label{Fig2}
\end{figure}

Complementary to the temporal analysis, Fig.~\ref{Fig2} illustrates the condensate density in real space and momentum space for a representative case with $F=1$ and $p(0)=\pi/2$. During the first Bloch period, $T \sim 6.29$, the density follows a well-defined oscillatory trajectory, and the momentum distribution remains organized by the driven quasimomentum dynamics. At longer times, this coherent pattern degrades; the real-space density broadens and fragments, while the momentum-space distribution develops multiple components [Fig.~\ref{Fig2}(a2),(b2)]. These features provide the field-driven manifestation of the same nonlinear dephasing mechanism identified in the collective-variable dynamics, consistent with the general picture of interaction-induced suppression of BO reported in Ref.~\cite{Anderson1998}. 

In conclusion, the numerical estimates in this subsection provide a quantitative bridge between the force-free phase-modulation mechanism and the driven BO dynamics. The preceding subsections identified three essential ingredients: nonlinear chirp generation through the internal phase-gradient drive, sign-sensitive feedback through the effective product \(\delta\mathcal{G}[\alpha]\), and sign-insensitive mobility suppression  through the factor \(e^{-\eta}\). Here, the explicit coefficients show how  these ingredients are realized for the present parameters.

The external force does not by itself damp BO. Its role is to sweep the quasimomentum periodically, thereby modulating the lattice-dispersion contribution through \(\cos(p_0-Ft)\). Since this Bloch-periodic factor has a bounded amplitude, the strength and direction of the chirp response are dictated by the nonlinear prefactors attached to it. Consequently, the suppression of BO is not produced by the sinusoidal modulation alone, but by its competition with the nonlinear bias set by the mean-field and LHY balance. The LHY correction reverses and filters the chirp-generation route, changes the effective feedback coefficient $\mathcal{G}[\alpha]$, and allows the accumulated magnitude of the chirp to suppress the mobility through $e^{-\eta}$.

Thus, the driven dynamics confirms the physical picture developed in the force-free analysis. The coherent BO is suppressed not by thermodynamic dissipation, but by a nonlinear transfer of center-of-mass motion into internal phase deformation. In this sense, the LHY correction acts as a nonlinear phase-response modifier; it reshapes the balance between nonlinear chirp generation and Bloch-periodic lattice forcing, and thereby
converts an initially coherent BO into a progressively arrested trajectory.

%

%
%

\section{Shallow Lattices}

\subsection{Nonlinear Spectral Structure}
While the previous section analyzed the tight-binding dynamics using the static tilted-lattice representation, we now examine the shallow-lattice regime for $V_0 \leq 1$. In this limit, where the wavepacket is no longer well-localized on discrete sites, the physics of Bragg reflection and interband tunneling is more transparently described in momentum space. As established in Sec.~II, the lattice acceleration $\alpha_L$ can be mapped into the co-moving frame. By applying the corresponding unitary transformation, the real-space evolution is governed by the GPE in the accelerating frame: 
\begin{align}
i\frac{\partial \Psi}{\partial t} = -\frac{1}{2} \left(\frac{\partial}{\partial x} + i\alpha_{L} t\right)^2 \Psi + V_0\sin^2\left(\frac{\pi x}{2}\right)\Psi + c_L|\Psi|^2 \Psi - d_L|\Psi|\Psi.  \label{Eq-accel-gp}
\end{align} 

While the dynamics near the zone edge $p_0 = \pi/2$ are dominated by the Bragg reflection between the two near-resonant momentum states, the wavefunction can be simplified using a two-level superposition of plane waves:
\begin{align}
\Psi(x,t) = c_a(t) e^{ikx} + c_b(t) e^{i(k-\pi)x}, \label{Eq-wf-2level}
\end{align}  
where $c_a(t)$ and $c_b(t)$ ensure particle conservation. 
Projecting Eq.~(\ref{Eq-accel-gp}) onto this basis reduces the infinite-dimensional GPE to an effective $2\times 2$ Hamiltonian in a minimal coupling approximation as 
\begin{align}
H(v_T) = \frac{1}{2} \left(
\begin{array}{cc} 
        v_T + s_{ML}Q & -V_0/2 \\
        -V_0/2 &  -v_T - s_{ML}Q 
    \end{array}
\right), \label{Eq-HvT}
\end{align}
where $v_T = \alpha_L t$ denotes the time-dependent detuning, $Q = |c_b|^2 - |c_a|^2$ is the population imbalance, and $s_{M L} = c_L - 3 d_L/4$ is the effective nonlinear parameter. This term introduces a nonlinear self-detuning that dynamically modifies the tunneling gap near $v_T+s_{M L} Q=0$.

By writing $c_\xi=\xi e^{i\theta_\xi}$, with $\xi=a,b$, and imposing the phase-locking condition
$\sin(\theta_b-\theta_a) = 0$, the nonlinear eigenvalue problem generated by Eq.~(\ref{Eq-HvT}) can be reduced to a self-consistent dispersion equation for the eigenenergy $\mu$. Since the population imbalance enters the Hamiltonian via nonlinear detuning, it must be determined self-consistently with the eigenvector. Eliminating this variable yields a quartic equation
\begin{align}
16 \mu^4 + 16 s_{ML} \mu^3
+ \left(4s_{ML}^2 - V_0^2 - 4v_T^2\right)\mu^2
- s_{ML}V_0^2 \mu
- {s_{ML}^2 V_0^2}/{4}
=0 .
\label{Eq-E4final}
\end{align}
The roots of Eq.~(\ref{Eq-E4final}) determine the nonlinear spectral structure as a function of the sweep parameter $v_T$. This self-consistency is essential. If one only enforces phase compatibility and occupation positivity, $a^2\ge0$ and $b^2\ge0$, the calculation merely gives a necessary physicality condition, $(s_{ML}-v_T)^2 \leq 4\mu^2- {V_0^2}/{4} \leq (s_{ML}+v_T)^2$, without determining the full set of stationary nonlinear eigenstates. Such a reduction treats the population imbalance as an amplitude constraint, thereby ignoring the feedback mechanism. Conversely, the quartic equation above restores this feedback and therefore captures the multivalued stationary branches. Consequently, unlike the linear Bloch bands, the nonlinear self-detuning controlled by $s_{ML}$ can generate nontrivial spectra, which provide the basis for the subsequent phase-space analysis.

\subsection{Looped Bloch Bands} 

\begin{figure*}[t!]
\includegraphics[width=0.9\textwidth]{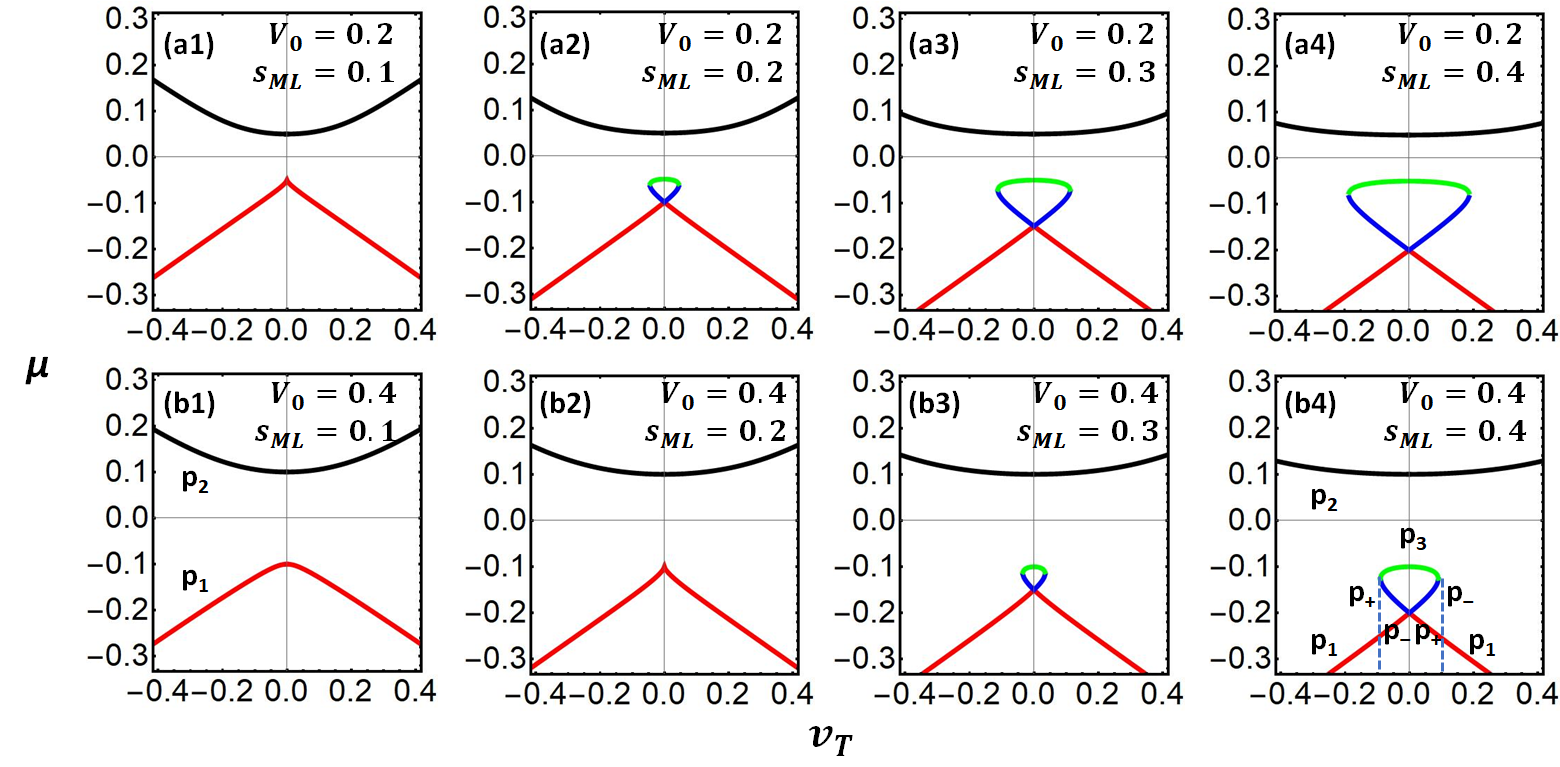}
\caption{(color online) Adiabatic energy levels $\mu$ as a function of the sweep parameter $v_T$ for $s_{ML}=0.1, 0.2, 0.3$, and $0.4$. Panels (a1)-(a4) correspond to a lattice depth of $V_0=0.2$, while panels (b1)-(b4) correspond to $V_0=0.4$. A loop structure emerges at the zone edge when the nonlinearity exceeds the critical threshold $s_{ML} > V_0/2$. The eigenstates are represented by the fixed points $p_i$ of the classical Hamiltonian $H_J$. In the multivalued regime, $p_3$ (green curves) acts as an unstable saddle point on the phase diagram [Fig.~\ref{Fig4}], whereas the lower branch $p_1$, the upper branch $p_2$, and side branches $p_{\pm}$ are stable points. 
}\label{Fig3}
\end{figure*}

The adiabatic energy levels obtained from Eq.~(\ref{Eq-E4final}) are shown in Fig.~\ref{Fig3} for representative nonlinear regimes $s_{ML}= 0.1$ to $0.4$ and two lattice depths: $V_0 = 0.2$ [panels (a1)-(a4)] and $V_0 = 0.4$ [panels (b1)-(b4)]. The nonlinear self-detuning reshapes the ordinary avoided crossing by generating additional stationary branches. As the sweep parameter varies slowly, the spectrum evolves from a simple two-branch avoided crossing into a multivalued structure. Specifically, while only two real roots exist for $s_{ML} < V_0/2$, four real roots emerge once the nonlinearity exceeds the threshold $s_{ML} > V_0/2$, leading to a loop structure at the top of the lower energy band.

The loop width is specified by the extrema of the energy curve. By rearranging the quartic dispersion relation as $v_T^2 = (4\mu^2 - V_0^2/4)(2\mu + s_{ML})^2/4\mu^2$, the boundaries of the multivalued region can be identified via the turning-point condition $d v_T^2/d\mu = 0$. This gives the nontrivial critical energy $\mu_c = -(V_0^2 s_{ML}/4)^{1/3}/2$ and the corresponding critical sweep parameter 
\begin{align}
v_c = \left(s_{ML}^{2/3} - (V_0/2)^{2/3}\right)^{3/2}.
\end{align}
For $s_{ML}= 0.4$, this analytical boundary yields $v_c = \pm 0.19$ for $V_0=0.2$ [Fig.~\ref{Fig3}(a4)] and $v_c = \pm 0.09$ for $V_0=0.4$ [Fig.~\ref{Fig3}(b4)], respectively, perfectly capturing the onset of the multivalued spectral branches.

The same nonlinear reshaping can also be identified from the population imbalance. To make this connection explicit, we apply the Hellmann--Feynman theorem to the effective detuning $h = v_T + s_{ML} Q$. For the two instantaneous eigenbranches, one obtains
\begin{align}
\langle \partial H/ \partial h \rangle = \partial \mu_{\pm}/ \partial h = -Q/2,   
\end{align} 
and the general two-level eigenenergies for a fixed $Q$. Once the self-consistency condition $Q = \mp \big[{v_T + s_{ML} Q}\big]\times \left[{{\left(v_T + s_{ML} Q\right)^2+\left(V_0 / 2\right)^2}}\right]^{-1/2}$ for the imbalance is satisfied, this relation can be equivalently written as $\left(1-Q^2\right)$ $\times\left(v_T+s_{ML} Q\right)^2 = {V_0}^2 Q^2 / 4$, yielding a quartic equation in $Q$ as a function of the sweep parameter $v_T$. This demonstrates that the loop structure features not only a multivalued energy spectrum but also a nonlinearity-mediated, multivalued population-imbalance response.

To understand the origin of this loop, we map the two-level GPE onto a classical Josephson-analog Hamiltonian in terms of the population imbalance $Q=|c_b|^2-|c_a|^2$ and the relative phase $\theta = \theta_b-\theta_a$: 
\begin{align}
H_J(Q,\theta) = v_T Q + \frac{1}{2}s_{ML} Q^2 + \frac{V_0}{2}\sqrt{1-Q^2}\,\cos\theta.   
\end{align}
The variables $Q$ and $\theta$ form a canonical conjugate pair, yielding the stationary-state conditions
\begin{align}
\frac{dQ}{dt} &= -\frac{\partial H_J}{\partial \theta} =  \frac{V_0}{2} \sqrt{1-Q^2}\,\sin\theta = 0, \label{Eq-dQdt}\\
\frac{d \theta}{dt} &= \frac{\partial H_J}{\partial Q} = v_T + s_{ML}Q  - \frac{ V_0 Q \cos\theta}{2 \sqrt{1 - Q^2}} = 0. \label{Eq-dtheta}
\end{align} 
The former implies the phase-locked branches $\theta=0$ and $\theta=\pi$, which reduce Eq.~(\ref{Eq-dtheta}) to a quartic equation for the equilibrium population imbalance:
\begin{align}
Q^4 + \frac{2 v_T}{s_{ML}}Q^3 + \left(\frac{v_T^2}{s_{ML}^2} + \frac{V_0^2}{4 s_{ML}^2} - 1\right)Q^2 - \frac{2v_T}{s_{ML}}Q - \frac{v_T^2}{s_{ML}^2} = 0, \label{Eq-Q4}
\end{align}
This equation is identical to the self-consistency condition obtained via the Hellmann--Feynman theorem and is structurally equivalent to the quartic energy dispersion in Eq.~(\ref{Eq-E4final}), reinforcing that the loop originates from the nonlinear feedback of the population imbalance. Within the interval $-v_c<v_T<v_c$, Eq.~(\ref{Eq-Q4}) admits four real roots, corresponding to the four stationary branches in the looped spectrum. A discriminant analysis of this quartic equation (detailed in Appendix~\ref{App:shallow-1}) yields $v_c = \left(s_{ML}^{2/3} - (V_0/2)^{2/3}\right)^{3/2}$, in agreement with the turning-point condition derived directly from Eq.~(\ref{Eq-E4final}).

Rather than just reproducing the loop boundary, the Josephson representation elevates the population imbalance and relative phase to canonical conjugate variables, thereby framing the multivalued stationary solutions as fixed points of $H_J(Q,\theta)$. This establishes a rigorous foundation for the ensuing phase-space analysis, where the stability of these fixed points, the emergence of separatrices, and the corresponding action variations reveal how the looped spectrum controls nonlinear LZ tunneling.

\begin{figure*}[h!]
\includegraphics[width=0.88\textwidth]{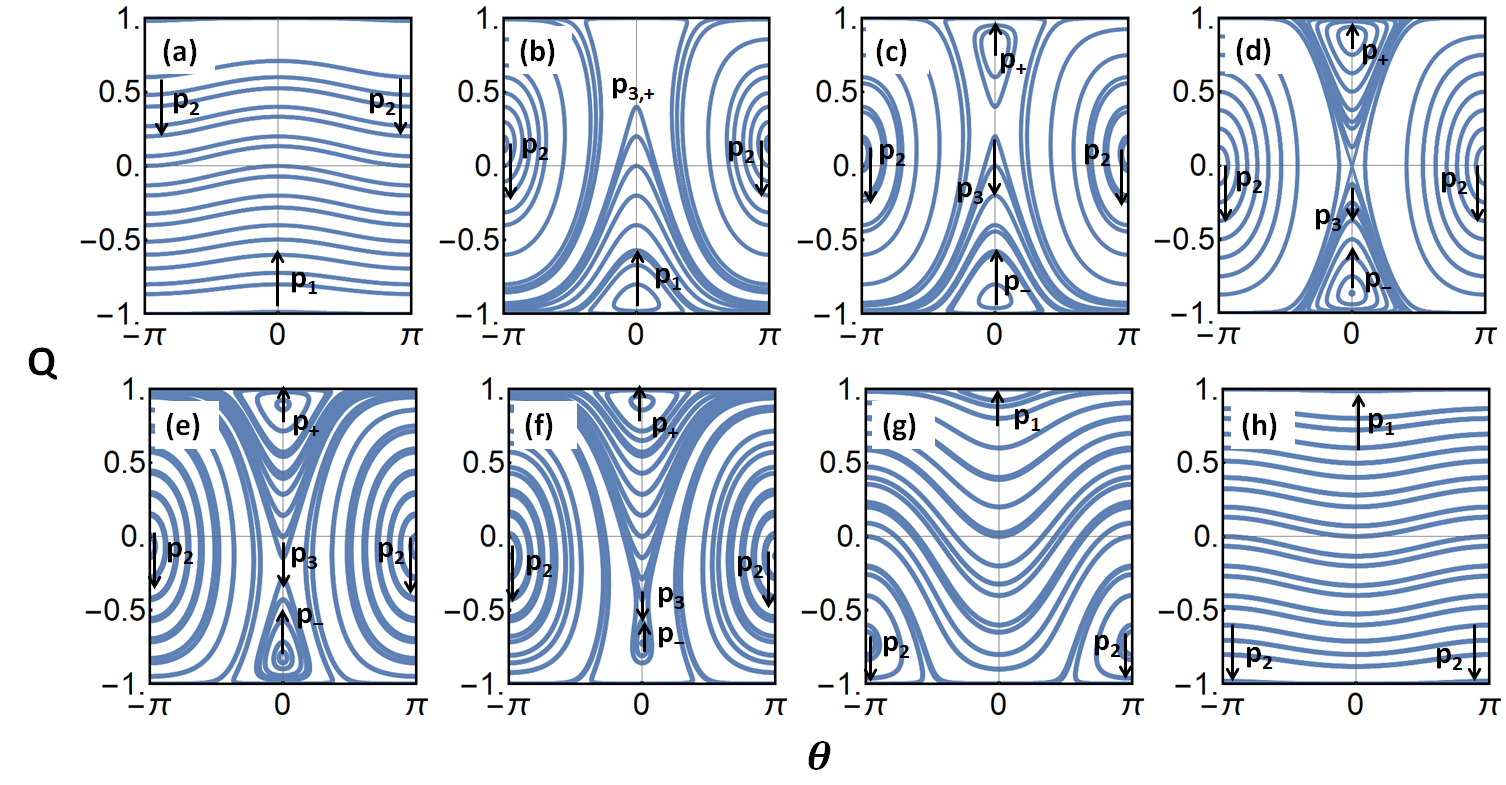}
\caption{(color online) Phase-space trajectories $(Q, \theta)$ for representative sweep parameters: (a) $v_T = -3$, (b) $v_T = -v_c = -0.09$, (c) $v_T = -0.05$, (d) $v_T = -0.009$, (e) $v_T = 0.04$, (f) $v_T = 0.08$, (g) $v_T = 0.5$, and (h) $v_T = 3$. Parameters correspond to $V_0=0.4$ and $s_{ML}= 0.4$ as in Fig.~\ref{Fig3}(b4). The arrows indicate the continuation of the fixed points as $v_T$ increases. The loop window is bounded by two saddle-node events: the creation of $p_3$ and $p_+$ at $v_T = -v_c$, and the annihilation of $p_3$ and $p_-$ at $v_T = +v_c$. Between these boundaries, the saddle $p_3$ generates a separatrix enclosing finite phase-space islands; the variation of their total area corresponds to a change in the classical action, providing the phase-space origin of nonlinear LZ tunneling.
}\label{Fig4}
\end{figure*}

\subsection{Classical Dynamics and Phase-Space Trajectory}

While the looped structure in the nonlinear Bloch bands reflects the modification of the adiabatic spectrum, its dynamical consequences are encoded in the phase-space evolution. In particular, the area enclosed by the loop is directly tied to the LZ tunneling probability \cite{Wu2003, Liu2002}.

To expose this connection, we analyze the phase-space trajectories ${\mathfrak z}(t)=(Q(t),\theta(t))$ for representative sweep parameters, as shown in Fig.~\ref{Fig4}(a)--(h), using $V_0=0.4$ and $s_{ML}=0.4$ as in Fig.~\ref{Fig3}(b4). For each fixed value of $v_T$, trajectories correspond to the isoenergetic contours of the Josephson Hamiltonian $H_J(Q,\theta)$, along which the Hamiltonian flow is governed by the equations of motion Eqs.~(\ref{Eq-dQdt}) and (\ref{Eq-dtheta}).
The stability of each fixed point is determined by the local curvature of $H_J$. A local extremum of $H_J$ in the reduced phase space is encircled by closed isoenergetic contours, corresponding to an elliptic fixed point with bounded oscillatory motion. By contrast, a saddle of $H_J$ exhibits opposing curvatures along the canonical directions; the separatrix emerging from the hyperbolic point divides the phase space into dynamically distinct regions. This topology criterion allows us to identify $p_1$, $p_2$, and $p_{\pm}$ as stable elliptic centers, and $p_3$ as a hyperbolic saddle point.

To understand the rearrangement of the stationary phase-space structure as the sweep parameter $v_T$ varies, we track the continuation of the fixed points by evaluating 
\begin{align}
\frac{dQ}{dv_T} = -\left[s_{ML} - \frac{V_0\sigma}{2(1-Q^2)^{3/2}}\right]^{-1},
\end{align}
where $\sigma = \cos\theta =\pm 1$. 
%
For $V_0=0.4$ and $s_{ML}=0.4$, the $\theta=\pi$ branch with
$\sigma = -1$ maintains a strictly positive denominator, dictating that
$dQ/dv_T < 0$. Consequently, the fixed point on this branch moves toward smaller $Q$ as $v_T$ increases. Conversely, on the $\theta=0$ branch, where $\sigma = 1$,  the denominator can change sign or vanish near a saddle point. The elliptic fixed points on this branch move toward larger $Q$ ($dQ/dv_T > 0$), whereas the intermediate hyperbolic fixed point moves toward smaller $Q$ ($dQ/dv_T < 0$). As $v_T$ approaches the edge of the loop window, this saddle point collides and annihilates with one of the elliptic centers via a saddle-node bifurcation. The arrows in Fig.~\ref{Fig4} illustrate the parameter-driven evolution. 
Crucially, this continuation does not represent a dynamical trajectory along an isoenergetic contour, but rather a structural restructuring of the phase space. The singularity in $dQ/dv_T$ marks the creation and annihilation of these fixed points, providing a clear phase-space origin for the multi-valued stationary solutions observed within the loop window.

The correspondence between the spectral branches and the phase space fixed points can be determined directly from the phase-locking structure of the two-level Hamiltonian $H(v_T)$. From the eigenvalue equation and the self-consistency condition, the amplitude ratio is explicitly related to the energy and population imbalance via ${c_b}/{c_a} = -{4\mu(1 + Q)}/{V_0}$. Since $Q > -1$ holds physically, the sign of the ratio dictates the phase-locked angle. Thus, a positive-energy branch with $c_b/c_a < 0$ corresponds to $\theta=\pi$, whereas a negative-energy branch with $c_b/c_a > 0$ belongs to the $\theta=0$ manifold.
This classification explicitly connects the color-coded energy dispersion in Fig.~\ref{Fig3} and the fixed points in Fig.~\ref{Fig4}. The black upper band, for which $\mu > 0$, maps onto the elliptic center $p_2$ on the $\theta=\pi$ branch. The negative-energy branches belong to the $\theta=0$ manifold. Within the loop window, the green branch corresponds to the hyperbolic saddle point $p_3$, while the red and blue branches correspond to the two elliptic fixed points denoted by $p_{\pm}$ according to the sign of $Q$. Outside the loop window, only one regular $\theta=0$ fixed point remains, which we denote by $p_1$.

At large negative bias [Fig.~\ref{Fig4}(a)], the tilt term $v_T Q$ dominates $H_J$, thus the phase space consists of open trajectories, indicating that the system remains adiabatically locked within a single band \cite{Pratse2024}. As $v_T$ increases, the fixed points $p_1$ and $p_2$ evolve along $\theta=0$ and $\theta = \pi$, respectively. In the linear limit, these two branches are related by the mirror relation $Q_{\theta = \pi}(v_T) = - Q_{\theta = 0}(v_T)$. The nonlinear self-detuning term $s_{\rm ML}Q$ breaks this simple branch symmetry, deforming the fixed-point structure and giving rise to multiple stationary solutions and multistability \cite{Gui2024}.

Within the loop window, $-v_c< v_T < v_c$ [Fig.~\ref{Fig4}(c)--(f)], the phase-space structure is no longer characterized by a single elliptic fixed point on the $\theta=0$ branch. Instead, three fixed points coexist on this branch: two elliptic centers, denoted by $p_-$ and $p_+$ according to the sign of $Q$, and one hyperbolic saddle point $p_3$. The centers $p_-$ and $p_+$ are encircled by closed orbits in the $Q<0$ and $Q>0$ regions, respectively. The saddle $p_3$ generates the separatrix that partitions these oscillatory regions from the running-phase trajectories, analogous to the macroscopic quantum self-trapping in coupled condensates \cite{Smerzi1997, Albiez2005}.

The boundaries of this window are defined by saddle-node bifurcations. For a sweep from negative to positive $v_T$, a saddle-center pair is created at $v_T = -v_c$ on the positive-$Q$ side. This pair gives rise to the saddle $p_3$ and the elliptic center $p_+$ [Fig.~\ref{Fig4}(b)]. As $v_T$ increases, $p_+$ moves toward larger $Q$, while $p_3$ moves downward across the $\theta=0$ branch. At $v_T = +v_c$, the saddle $p_3$ collides with the negative-$Q$ elliptic center $p_-$ and annihilates. This collision destroys the additional closed-orbit island, restoring the phase space to its two-branch structure at large positive bias [Fig.~\ref{Fig4}(g)--(h)]. The disappearance of the adiabatic fixed-point island provides the phase-space mechanism for population transfer across the nonlinear bands. Therefore, the loop window is physically a coexistence regime bounded by discrete saddle-node bifurcations, rather than a region hosting a single, continuous collision.

The separatrix enclosing these islands bounds a finite phase-space area, defining the classical action relevant to adiabatic transport. As the system is driven across the saddle-point boundary, the classical action exhibits a discontinuous jump, implying the topology change of the accessible trajectories and a finite probability for interband tunneling. Therefore, the separatrix signals the breakdown of adiabatic evolution. This provides a direct dynamical interpretation of nonlinear LZ tunneling in terms of phase-space topology: the tunneling probability is controlled not only by the instantaneous energy gap, but also by the separatrix structure generated by the looped band. A quantitative analysis of the stability of the fixed points in the phase-space dynamics is provided in Appendix \ref{App:shallow-1-phasedynamics}.

Critically, the phase-space evolution reveals the fixed-point counterpart of the looped spectrum in Fig.~\ref{Fig3}(b4). Rather than being a mere collection of alternative stationary branches, the loop represents a topological obstruction that breaks the continuous tracking of the initial adiabatic state. During the parametric sweep, the negative-$Q$ elliptic island governing the initial population character is progressively squeezed by the separatrix generated from $p_3$ and ultimately destabilized at the upper saddle-node bifurcation. Beyond this boundary, the only surviving $\theta=0$ elliptic branch carries positive population imbalance. Consequently, the nonlinear loop enforces an asymmetric adiabatic switch from a negative-$Q$ to a positive-$Q$ branch, providing the fixed-point interpretation of nonlinear state transfer.

\section{LHY-Suppressed Nonlinear LZ Tunneling}
 
In the shallow-lattice regime, the dynamics are naturally formulated in momentum space, where the accelerated quasimomentum drives the system across an avoided crossing. The preceding phase-space analysis identified the looped spectrum as a topological mechanism for nonlinear state transfer. In this section, we quantitatively present how the breakdown of adiabatic following appears as a nonadiabatic LZ transition. While linear LZ mechanism provides the baseline exponential scaling, the mean-field and LHY terms reshape the effective detuning and the phase-space topology, thus leading to an interaction-dependent tunneling probability beyond the conventional linear result.

\subsection{Nonadiabatic Transition at an Avoided Crossing}

\begin{figure}[h!]
   \includegraphics[scale=0.18]{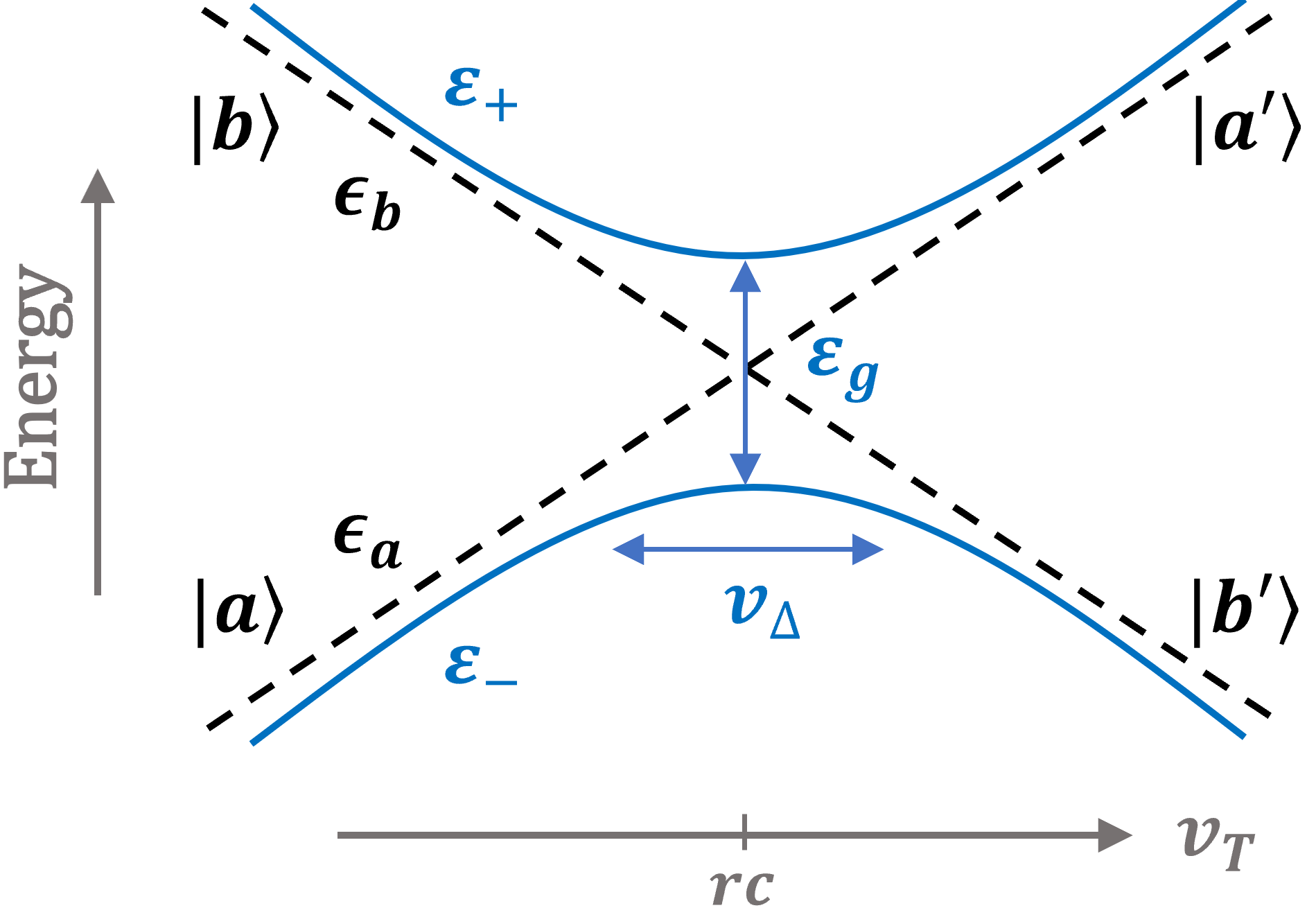}
   \caption{{Energy spectrum of the two-level system as a function of the sweep parameter $v_T$. The dashed lines denote the diagonal energies of the Hamiltonian without the optical lattices, while the solid lines represent the corresponding adiabatic eigenenergies, illustrating the avoided crossing structure.}} \label{Fig5}
 \end{figure}

We first consider the linear two-level avoided-crossing problem, which provides the reference LZ process before the mean-field and LHY-induced corrections are included. As illustrated in Fig.~\ref{Fig5}, the black dashed lines denote the diabatic energies $\epsilon_a(v_T)$ and $\epsilon_b(v_T)$ associated with the fixed basis channels. The states $|a\rangle$ and $|b\rangle$ label the incoming diabatic channels at $v_T<0$, while $|a'\rangle$ and $|b'\rangle$ denote the corresponding outgoing diabatic channels at $v_T>0$. In the absence of the off-diagonal optical-lattice coupling, these diabatic levels cross at the resonance center. When the coupling is included, the crossing is opened into an avoided crossing, and the blue solid curves represent the instantaneous adiabatic eigenenergies $\varepsilon_\pm(v_T)$. The minimal avoided crossing is characterized by its separation, $\varepsilon_g$, and its width $v_{\Delta}$ is evaluated at the resonance center ${rc}$ via $ v_{\Delta} \approx \varepsilon_{g} /\left({d |\epsilon_b-\epsilon_a|}/{d v_T}\right)_{rc}$. The width estimates the detuning interval over which the diabatic mixing is appreciable.

In the linear regime, the population transfer across this crossing can be described by the standard LZ mechanism. To connect the present model with the canonical LZ problem, we recast the linearized two-level equations into the Weber form by introducing the scaled variable $z = \alpha_L^{1/2} e^{-i\pi/4} t$ and the dimensionless parameter $\gamma = V_0^2/16\alpha_L$. The resulting equations admit the asymptotic solutions that determine the transition probability between the diabatic states, and promote the nonadiabatic transition probability in terms of the LZ formula
\begin{align}
\Gamma_{\mathrm{LZ}} = \exp\!\left(-\frac{\pi V_0^2}{8\alpha_L}\right). \label{Eq-Gammaq}
\end{align}
This formula captures the exponential suppression of tunneling in the adiabatic limit $\alpha_L \to 0$. The detailed derivation of this result including the asymptotic analysis of the Weber equations, is presented in Appendix~\ref{App:shallow-2-2latticerule}. This result provides the reference behavior against which the mean-field and LHY-induced corrections will be evaluated in the following subsections.

\subsection{Adiabatic Tunneling Due To Nonlinearity}

In the linear regime, the nonadiabatic transition arises because a finite sweep prevents perfect following of the instantaneous adiabatic eigenstate. However, in a nonlinear system, the adiabatic following can break down even in the quasi-static sweep, because the followed phase-space island disappears at the separatrix and the saddle-node, thereby interrupting the smooth continuation of the original state.

We therefore consider the quasi-static limit in which $v_T$ changes slowly compared with the intrinsic motion on a certain Josephson closed orbit of energy $E = H_J(Q, \theta; v_T)$ and fixed $v_T$ within a period $T(E, v_T)$. In this limit, a closed orbit evolves through a sequence of nearly frozen phase portraits, and its action, 
\begin{align}
I(E, v_T) = \frac{1}{2\pi} \oint\, Q(\theta; E, v_T)\, d\theta, \label{Eq-actionI}
\end{align}
remains conserved according to the adiabatic theorem, except when the trajectory approaches the separatrix. Therefore, the tunneling dynamics induced by nonlinearity can be analyzed using the method of adiabatic invariants.

As the evolving orbit reaches the separatrix, the adiabatic invariant must be matched to the action of the outgoing running-phase trajectory \cite{Liu2002}. Far from the separatrix, this trajectory eventually approaches a straight line in the phase space, where the final population imbalance becomes a constant $Q = Q_f$. In the sense of measuring the tunneling probability to the final occupation at $|a'\rangle$, the relevant action $I = 1- Q_f$ is the normalized phase-space area between the asymptotic horizontal orbit \(Q=Q_f\) and the population-imbalance boundary \(Q=1\).
%
Through the adiabatic-invariant matching, the final outgoing action is identical to the critical separatrix action $I_c$ evaluated at the upper saddle-node boundary, $v_T = +v_c$. 
This same quantity is twice the final population in the outgoing \(a\)-channel, corresponding to $|a'\rangle$ in the Fig.~\ref{Fig5}. 
Consequently, the adiabatic tunneling probability is given by 
\begin{align}
\Gamma_{ad} = 2 \pi |c_{a,f}|^2 = \pi {I_c}, \label{Eq-Gammaad}
\end{align}
where the saddle-node point lies on the phase-locked branch $\theta = 0$.

At this boundary, the critical population imbalance is the negative-branch solution of Eq.~(\ref{Eq-Q4}), $Q_c = -Q_s$, with 
\begin{align}
Q_{s} = \sqrt{1- \left(\frac{V_0}{2s_{ML}}\right)^{2/3}}. \label{Eq-Qs}
\end{align}
For the parameters of Fig.~\ref{Fig4}, \(V_0=0.4\) and \(s_{ML}=0.4\), this gives \(v_c\sim 0.090\) and \(Q_c\sim -0.608\). This value corresponds to the limiting collision point approached by \(p_-\) and \(p_3\) as Fig.~\ref{Fig4}(f) is continuous from \(v_T = 0.08\) to \(v_T = +v_c\). The critical energy then fixes the homoclinic trajectory through \(H_J(Q,\theta;v_c) = E_{c}\).
%
The phase-space area associated with this separatrix can then be analytically evaluated in the vicinity of the critical region defined by $\delta \equiv 2 s_{ML}/V_0 - 1 \to 0$, where the loop structure emerges. In this regime, the critical sweep parameter is approximated as
\begin{align}
v_c = \frac{V_0}{2} \Bigl[(1+\delta)^{2/3}-1 \Bigr]^{3/2} \simeq \frac{V_0}{2} \left(\frac{2 \delta}{3}\right)^{3/2}.
\end{align}
Within the quasi-static approximation, the Josephson energy is locally conserved near the separatrix, and a small deviation $\Delta Q$ from the critical point $Q_c$ can be introduced to characterize the width of the homoclinic orbit.

Although the saddle-node location is fixed by \(Q_c\), the area estimate requires the \(Q\)-extent of the homoclinic lobe. We denote the two relevant near-critical roots controlling this extent by \(Q_{\rm t}^{-}\) and \(Q_{\rm t}^{+}\), and define $\Delta Q = Q_{\rm t}^{+}-Q_{\rm t}^{-}$.
To the leading order in $\delta$, this yields the quantity
\begin{align}
\Delta Q \approx h_1(\varrho) + \frac{1}{2}\sqrt{h_2(\varrho)-\frac{2 v_c(1+\varrho^2)}{s_{ML} h_1(\varrho)}} + \frac{1}{2}\sqrt{h_2(\varrho)+\frac{2 v_c(1+\varrho^2)}{s_{ML} h_1(\varrho)}},
\end{align}
where $\varrho = V_0/2s_{ML}$, $h_1(\varrho) = \sqrt{1-\varrho^2+2\varrho^{4/3}-2\varrho^{3/2}}$, and $h_2(\varrho) = -1 +\varrho^2+8\varrho^{2/3}-8\varrho^{4/3}+3 v_c^2/s_{ML}^2$. 
In the strong nonlinearity regime, the leading-order expansion in $\delta$ simplifies this result to $\Delta Q \sim \sqrt{6\delta}$, which captures the dominant scaling behavior near the bifurcation point. Near the lower boundary \(Q_{\rm t}^{-}\) of the lobe and around \(\theta=0\), the corresponding phase-space angular width is
\begin{align}
\Delta\theta \simeq  4\sqrt{v_c/V_0}\,(Q-Q_t^-)^{1/2} + 2\sqrt{2 v_c/V_0}\,(Q-Q_t^-)^{3/2}.
\end{align}
Substituting this relation into the critical-area integral gives the tunneling probability
\begin{align}
\Gamma_{ad} = \frac{1}{4\pi}\int_{Q_t^-}^{Q_t^+}\, \Delta\theta\, dQ \simeq \frac{2}{3} \delta^{3/2}.
\end{align}
This result represents a small but finite adiabatic deviation generated by the nonlinear saddle-node separatrix, even in the quasi-static limit.

\subsection{Nonadiabatic Tunneling Near Critical Point}

%
When the sweep rate $\alpha_L$ is finite, the slowly varying bias \(v_T(t)\) becomes an explicit source of action variation, turning the adiabatic invariant \(I\) into a slowly varying dynamical variable. The nonadiabatic tunneling near the critical point can therefore be formulated by applying the Hamilton-Jacobi formalism to the frozen Josephson dynamics with Hamiltonian $H_J$. As mentioned previously for each fixed \(v_T\), the Josephson dynamics are integrable in the two-dimensional phase space \((Q,\theta)\), and its regular trajectories are periodic, including both librational closed orbits and rotational running-phase orbits. This periodicity allows one to introduce action-angle coordinates, with the action \(I\) in Eq.~(\ref{Eq-actionI}) chosen as the canonical momentum and its conjugate coordinate \(\phi\) as the angle variable.

For a fixed \(v_T\), the Hamiltonian \(H_J(Q, \theta; v_T)\) has no explicit time dependence. The canonical transformation from the original variables \((\theta,Q)\) to the action-angle variables \((\phi,I)\) can be characterized by the type-2 generating function $S(\theta, I; v_T)$, whose total differential is written as 
\begin{align}
dS = Q d\theta + \phi dI + (K_J - H_J)\,dt.
\end{align}
Here, $H_J$ denotes the Hamiltonian in the original variables, while $K_J$ denotes the transformed Hamiltonian in the action-angle variables. The canonical relations are then written as
\begin{align}
Q = {\partial S}/{\partial \theta}, 
\qquad
\phi= {\partial S}/{\partial I}, 
\end{align}
together with $K_J = H_J + \partial S/\partial t$. As there is no explicit time dependence in the generating function at fixed $v_T$, we have $\partial S/\partial t = 0$. The transformed Hamiltonian is therefore the original Josephson Hamiltonian expressed in action-angle variables. For a regular frozen orbit, the angle dependence is thereby removed, and the transformed Hamiltonian can be written as, \(K_J(I;v_T) = E(I;v_T)\), where $ E(I;v_T)$ denotes the energy of the corresponding frozen Josephson orbit.
Hamilton's equation then gives
\begin{align}
\dot{I} &= -\partial K_J/\partial \phi = 0, \nonumber\\
\dot{\phi} &= \partial K_J/\partial I = {\partial E(I;v_T)}/{\partial I}
\equiv \omega(I;v_T).
\end{align}
Thus, $\omega$ is the oscillation frequency of the fixed orbit, and \(\phi\) increases by \(2\pi\) over one orbital period. The action-angle variables therefore provide a direct route to the oscillation frequency without solving the full equations of motion \cite{Landau1976}.

One may describe the fixed canonical transformation by Landau's type-1 generating function $F\left(\theta, \phi ; v_T\right)$, for which $Q = \partial_\theta F$ and $I = -\partial_\phi F$. Its monodromy over one angle cycle gives $\Delta_\phi F = -2 \pi I$, intuitively demonstrating that the action directly represents the phase-space area associated with the multivalued part of $F$. However, this derivation does not simply lead to Eq.~(\ref{Eq-Gammaad}), since the tunneling probability still requires matching this critical action to the outgoing population area. As a result, the type-2 form $S\left(\theta, I ; v_T\right)$ is more convenient for the time-dependent sweep extension, because $I$ is kept as an independent variable and the perturbation from $\dot{v}_T$ follows directly from $\partial S / \partial v_T$.

When the sweep parameter is allowed to vary in time, $v_T = v_T(t)$, the same fixed generating function is used as an instantaneous generating function, $S(\theta, I; v_T) \rightarrow S(\theta, I; v_T(t))$. 
Its parametric dependence on \(v_T(t)\) produces an explicit time-dependent contribution to the transformed Hamiltonian, so that the action is no longer exactly conserved. The explicit time dependence enters through
${\partial S(\theta,I;v_T(t))}/{\partial t} =
\left({\partial S(\theta,I;v_T)}/{\partial v_T}\right)_{\theta,I}
\dot v_T(t) $. We define $\Lambda_0(\theta,I;v_T) = 
\left({\partial S(\theta,I;v_T)}/{\partial v_T}\right)_{\theta,I}$. After taking this derivative, the resulting function is expressed in the action-angle variables as $\Lambda(I,\phi;v_T) \equiv \Lambda_0(\theta(I,\phi;v_T),I;v_T)$. Here, $\theta=\theta(I,\phi;v_T)$ follows from the canonical transformation generated by $S$. Since \(\Lambda\) is a single-valued function on the frozen orbit, it is \(2\pi\)-periodic in \(\phi\). The effective Hamiltonian is therefore 
\begin{align}
K'_J(I, \phi; t) = E(I; v_T(t)) + \Lambda(I, \phi; v_T(t)) \dot{v}_T(t).
\end{align}
As previously defined, ${v}_T(t) = \alpha_L t$ represents the time-dependent detuning. In the linear approximation, the frequency chirp $\alpha_L = \dot{v}_T(t)$ determines the constant lattice acceleration. The canonical equations of motion then become
\begin{align}
\dot{I} &= -\frac{\partial K'_J}{\partial \phi} = -\dot{v}_T(t) \left(\frac{\partial \Lambda}{\partial \phi}\right)_{I, v_T}, \label{Eq-Idot}\\ 
\dot{\phi} &= \frac{\partial K'_J}{\partial I} = \omega(I; v_T(t)) + \dot{v}_T(t) \left(\frac{\partial \Lambda}{\partial I}\right)_{\phi, v_T}. \label{Eq-phidot}
\end{align}

In the vicinity of the fixed point $p_-$, the closed orbit shrinks to the fixed point and the enclosed phase-space area vanishes, so that $I \approx 0$. Linearizing the fixed Josephson dynamics Eqs.~(\ref{Eq-dQdt}) and (\ref{Eq-dtheta}) around $p_-$ by setting $Q = Q_-(v_T) + \varepsilon_Q$ and $\theta = 0 + \vartheta_\theta$, we obtain 
$\varepsilon_Q(t) = \varepsilon_{0} \cos\tilde{\omega} t$ and $\vartheta_\theta(t) = \vartheta_0 \sin\tilde{\omega} t$, 
with the local orbit frequency near the elliptic fixed point
\begin{align}
\tilde{\omega} = \left(\frac{V_0}{2}\right)\sqrt{\frac{1}{1-Q^2}-\frac{2s_{ML}}{V_0}\sqrt{1-Q^2}}. \label{Eq-tildeomega}
\end{align}
It should be noticed that the linearized motion and the action-angle description refer to the same small closed orbit, so that the population imbalance $Q$ in Eq.~(\ref{Eq-tildeomega}) is evaluated on the frozen branch $p_-$, where $Q = Q_-(v_T)$. Consequently, the relationship between the tunneling rate and the action reduces to the calculation of the enclosed area spanned by $\varepsilon_Q$ and $\vartheta_\theta$. Since $\omega\left(I ; v_T\right)= 2 \pi / T\left(I ; v_T\right)$, the local small-oscillation frequency gives the small-action limit of the action-angle frequency, $\omega\left(I ; v_T\right) \rightarrow \tilde{\omega}\left(v_T\right)$ as $I \rightarrow 0$. Thus, near $p_-$, the local frequency $\tilde{\omega}$ provides the leading approximation to $d\phi/dt$.
%
The real small-oscillation branch is bounded by the minimum imbalance magnitude $|Q|_{\min} = Q_s$ corresponding to \(Q = -Q_s\) on the \(p_-\) branch, where \(\tilde{\omega}\) vanishes. For \(|Q|>|Q|_{\min}\), the local frequency is real, whereas the analytic continuation into \(|Q|<|Q|_{\min}\) makes \(\tilde{\omega}\) pure imaginary. This boundary therefore provides the real precursor of the complex singular point used below.
%

%
%
We now consider the action change associated with the effective nonadiabatic transition during the sweep. The incoming and outgoing actions, $I_-$ and $I_+$, are defined in the asymptotic time regions before and after the LZ tunneling, where the action is again well defined with respect to the corresponding frozen Josephson dynamics. The time regimes are chosen such that the action variation outside them has become negligible, or contributes only through rapid oscillatory cancellation. Assume that the incoming trajectory is taken to follow the stable fixed-point branch before the transition, so that $I_- \approx 0$. The net change of the action can be written as \cite{Landau1976}
\begin{align}
\Delta I = I_+ - I_- = -\int_{-\infty}^{\infty} \left(\frac{\partial \Lambda}{\partial \phi}\right)\dot{v}_T \,dt.
\end{align}
To evaluate the change in action, we express \(\Lambda(I,\phi;v_T)\) in terms of its Fourier components,
\begin{align}
\Lambda(I, \phi; v_T) = \sum_{\ell=-\infty}^{\infty} e^{i\ell\phi}\Lambda_{\ell}(I; v_T), \qquad \Lambda_{-\ell}(I; v_T) = \Lambda_{\ell}^*(I; v_T). 
\end{align}
In the time-dependent problem, $\Lambda$ is evaluated along the trajectory $(I(t), \phi(t); v_T(t))$. When \(\phi(t)\) varies monotonically, the action change is thus
\begin{align}
\Delta I = -\int_{C_\phi} \frac{\partial \Lambda}{\partial \phi}\frac{d v_T}{dt}\frac{d t}{d\phi}\, d\phi.
\end{align}

For a slow sweep, the amplitudes $\dot{v}_T(t)$ and $\Lambda_{\ell}\left(I ; v_T(t)\right)$ vary slowly, whereas the Fourier factors $e^{i \ell \phi(t)}$ oscillate rapidly because, to leading order in the sweep rate, $\dot{\phi} \simeq \omega\left(I ; v_T(t)\right)$ and many orbital cycles are completed during the time scale over which $v_T(t)$ changes appreciably. The real-time integral for $\Delta I$ is therefore highly oscillatory, and its leading exponential contribution is controlled by the closest nonanalytic point of the analytically continued integrand.
By continuing the phase variable \(\phi\) into the complex plane, this point occurs at \(\phi=\phi_0\), where the local small-oscillation frequency \(\tilde{\omega}\) vanishes. Equivalently, since \(d\phi/dt\simeq\tilde{\omega}\) near \(p_-\), the factor \(dt/d\phi\) becomes singular at this point. The corresponding complex phase is given by
\begin{align}
\phi_0 &= \int^{t_s}_{t_r} \omega(I, v_T(t))\,dt 
= \int_0^{{Q}_s} \frac{d\phi}{d t} \frac{dt}{d{Q}}\, d{Q} \nonumber\\
&= \left(\frac{V_0^2}{4\alpha_L}\right) \int_0^{{Q}_s} \left(1-{Q}^2 \right)^{1/4}
\left[\frac{1}{\left(1-{Q}^2\right)^{3/2}}-\frac{2s_{ML}}{V_0} \right]^{3/2} \,d{Q}.
\end{align}
With the form defined in Eq.~(\ref{Eq-Qs}), $Q_s$ here denotes the upper endpoint of the imaginary-frequency interval in the analytically continued coordinate. In the interval \(0< Q< Q_s\), the bracket in the last integral is negative, and the analytic branch for which the condition \({\rm Im}\,\phi_0 > 0\) holds is chosen.

Within this framework, the action increment is exponentially suppressed and takes the form
\begin{align}
|\Delta I| \simeq e^{-\ell\, {\rm Im}\,\phi_0}.
\end{align}
Near the fixed point $p_-$, the phase-space trajectories are approximately elliptical and correspond to a simple harmonic motion. In this regime, the local Hamiltonian takes the form $H_{\rm shm}(I;\omega) = I\omega$, with the canonical variables $p = -\sqrt{2M I \omega}\,\sin\phi$ and $q = \sqrt{{2I}/{M \omega}}\,\cos\phi$. The slow dependence on the sweep parameter enters through the local frequency \(\omega=\omega[v_T(t)]\). Therefore, the time-dependent correction satisfies $\Lambda \dot{v}_T = \Lambda_\omega \dot{\omega}$, where $\Lambda_\omega \equiv {\partial W}/{\partial\omega}$.  
The harmonic correspondence leads to 
\begin{align}
\Lambda_\omega = -\frac{I}{2\omega}\sin 2\phi.
\end{align}
This perturbation contains only the \(\ell=\pm2\) Fourier components, so the dominant exponential factor is governed by \(\ell=2\). The corresponding local equations of motion are
\begin{align}
\dot{I} = \frac{I}{\omega}\dot{\omega}\,\cos 2\phi, \qquad 
\dot{\phi} = \omega - \frac{\dot{\omega}}{2\omega}\,\sin 2\phi.
\end{align}
Identifying the exponentially small action increment with the nonlinear correction to the LZ transition probability, the tunneling probability retains an LZ-like form,
\begin{align}
\Gamma \sim e^{-\kappa\,{\pi V_0^2}/{8 \alpha_L}}, \label{Eq-Gammakappa}
\end{align}
with the weighting coefficient
\begin{align}
\kappa = \frac{4}{\pi} \int_0^{{Q}_s} \left(1-{Q}^2\right)^{1/4}
\left[\frac{2s_{ML}}{V_0}-\frac{1}{\left(1-{Q}^2\right)^{3/2}}\right]^{3/2} \,d{Q}.
\end{align}
Here, the expression in the square bracket is positive because $\kappa$ is defined from the positive real integral given by ${\rm Im}\,\phi_0$. The integrand is therefore real and positive over $0< Q< Q_s$.

\begin{figure}[t!]
\includegraphics[scale=0.6]{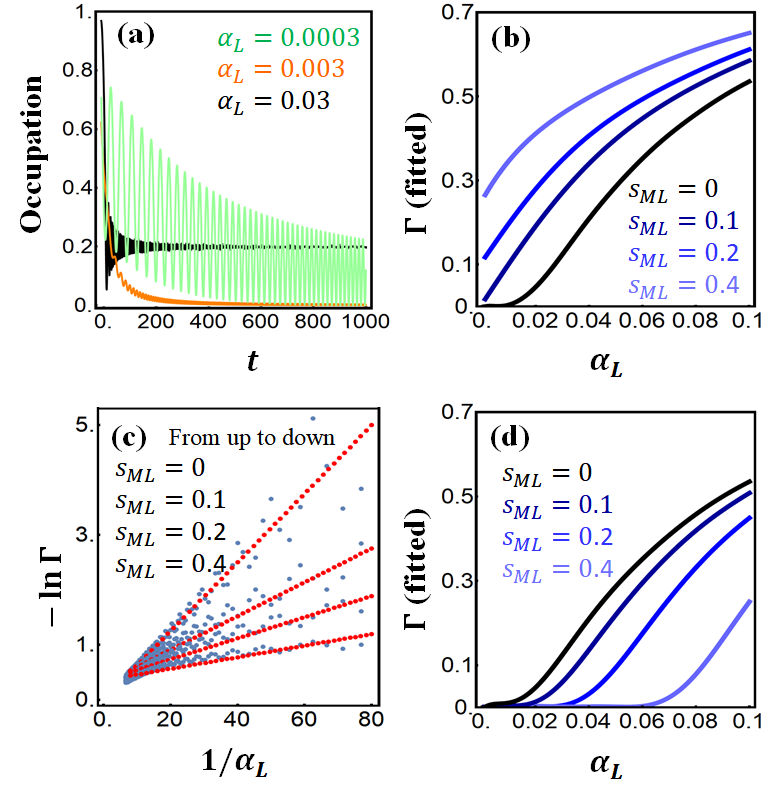}
\caption{(Color online) LZ dynamics of a quantum droplet initially prepared in the lower-band state $|a\rangle$. 
(a) Occupation probability $|c_a(t)|^2$ for driving strengths $\alpha_L = 0.0003,\ 0.003$, and $0.03$ at a fixed interaction strength $s_{ML} = 0$. At long times and for positive sweep velocity $v_T > 0$, $|c_a(t)|^2$ reflects the excitation probability resulting from adiabatic or nonadiabatic tunneling. 
(b) Extracted transition probability for increasing $s_{ML}$, illustrating a crossover governed by the competing effects of mean-field enhancement and LHY suppression. 
(c) Semi-log plots of $-\ln \Gamma$ vs $1/\alpha_L$ for the same $s_{ML}$ values, with effective weighting factors $\kappa$ extracted from the slopes. 
(d) Nonreciprocal tunneling from the upper to the lower band under the same interactions. The original exponential model fails to fit the saturation behavior at strong $s_{ML}$, whereas a modified expression incorporating a generalized exponent and prefactor captures the nonlinear response across different sweep-rate regimes.
}
\label{Fig6}
\end{figure}

To validate the analytical prediction of the tunneling probability, we compute the real-time dynamics of the droplet using the time-dependent two-level equation. Figure~\ref{Fig6}(a) shows the evolution of the occupation probability $|c_a(t)|^2$ for different driving strengths $\alpha_L=0.0003,\, 0.003$, and $0.03$ at fixed interaction strength $s_{ML}=0$.

In the adiabatic regime of small $\alpha_L$, the system exhibits long-lived BO before reaching the asymptotic state. These oscillations reflect the slow tunneling dynamics and lead to significant temporal fluctuations in the occupation probability, making the extraction of a well-defined transition rate nontrivial. This behavior is evident in Fig.~\ref{Fig6}(c), where the semi-log plots of $-\ln \Gamma$ versus $1/\alpha_L$ display noticeable scattering in the small-$\alpha_L$ regime. As the sweep rate increases, the dynamics cross over to a nonadiabatic regime, where the oscillations become damped and the transition probability stabilizes, consistent with the expected LZ behavior.

The dependence of the tunneling probability on the interaction strength is summarized in Fig.~\ref{Fig6}(b). The fitted curves for $s_{ML}=0,\,0.1,\,0.2$, and $0.4$ demonstrate a clear modification of the exponential scaling, reflecting the competing effects of the mean-field repulsion and the LHY-induced attraction, as well as the dynamical suppression by the latter. 
In particular, the extracted slopes in Fig.~\ref{Fig6}(c) yield effective weighting factors $\kappa=0.993,\,0.491,\,0.305$, and $0.167$, respectively, indicating a progressive reduction of tunneling sensitivity to the sweep rate as the nonlinear interaction becomes stronger. This trend is in direct agreement with the analytical prediction of the renormalized exponential factor in Eq.~(\ref{Eq-Gammakappa}).

A qualitatively different behavior emerges in the reverse process, where the system is initialized in the upper band. As shown in Fig.~\ref{Fig6}(d), the transition to the lower band is strongly suppressed by the nonlinearity, leading to a pronounced asymmetry between forward and backward tunneling \cite{Jona-Lasinio2003, Kitamura2020}. A very important signature of this asymmetry is the nonlinearly induced inertia against system variation, evidently showing again the localization tendency within the LHY field.      

To quantify this effect, we first apply the conventional exponential form, which assumes a simple inverse scaling with $\alpha_L$. While this model captures the qualitative suppression of tunneling at small $s_{ML}$, it fails to reproduce the plateau structure observed at the larger interaction strengths ($s_{ML} \gtrsim 0.2$), indicating a breakdown of the standard scaling. To account for this deviation, we introduce a generalized fitting form,
\begin{align}
\Gamma = A \exp \left[-\left(\frac{\pi V_0^2}{8 \alpha_L^n}\right)\left(1+\frac{\beta s_{ML}}{2 V_0}\right)\right],
\end{align}
which incorporates both a modified exponent $n$ and a prefactor $A$. For practical application, we obtain $A=0.944$, $\beta=-3.537$, and $n=1.245$ for $s_{ML}=0.1$, $A=0.821$, $\beta=-3.452$, and $n=1.845$ for $s_{ML}=0.2$, and  $A=0.757$, $\beta=-1.979$, and $n=3.214$ for $s_{ML}=0.4$. In addition to a fine agreement with numerical data for all $s_{M L}$, the modified function captures the broader shoulders and soft onset in the tunneling response. The exponent parameter $n>1$ suggests a tunneling suppression at a rate slower than the inverse of $\alpha_L$, revealing nonlinear and collective effects beyond simple perturbative scaling.

Taken together, the results presented in this section establish a unified picture of LZ tunneling for an accelerating droplet in a shallow lattice, where the nonlinearities originating from the beyond-mean-field interactions play a central role. In the adiabatic regime, the tunneling process is governed by the breakdown of action conservation at the separatrix, where the phase-space topology determines the transfer probability through the enclosed area. As the system enters the nonadiabatic regime, the dynamics are controlled by the complex-phase structure of the action, leading to an exponentially suppressed transition probability whose effective exponent is renormalized by the nonlinear interaction. 

The emergence of the weighting factor $\kappa$ reflects the modification of the classical trajectory in the complex plane, providing a direct link between the microscopic interaction parameters and the tunneling dynamics. While this renormalized exponential form captures the behavior within the weak-to-intermediate nonlinear regime, the numerical results reveal systematic deviations at stronger interactions, where the transition probability exhibits a broadened crossover and a scaling with the sweep rate that relaxes slower than a simple reciprocal dependence. This breakdown of the conventional LZ scaling signals the onset of collective effects beyond the perturbative framework and necessitates a generalized description incorporating both modified exponents and prefactors.

Overall, the interplay between the mean-field repulsion and the LHY-induced attraction tailors both the phase-space structure and the nonadiabatic response via competition between enhancement and suppression, leading to a rich tunneling behavior that interpolates between classical adiabatic transport and strongly nonlinear, nonperturbative dynamics.

\section{Conclusion}
We have developed a unified description of the nonlinear Bloch dynamics and LZ tunneling of quantum droplets in one-dimensional optical lattices, bridging the deep- and shallow-lattice regimes within a common phase-space framework. In the deep-lattice limit, the tight-binding description captures how the mean-field and the LHY interactions renormalize the band structure, enabling self-trapping, breathing, and BO via the dispersion suppression and the phase modulation. In the shallow-lattice regime, the dynamics are naturally formulated in momentum space, where acceleration drives the system across an avoided crossing and reveals the interplay between the nonlinear spectral deformation and the nonadiabatic transitions.

The force-free limits show that the key deep-lattice mechanism is a chirp-mediated transfer of coherent transport into internal deformation. When the conserved quasimomentum is initialized at $p_0=0$, the LHY correction activates an internal breathing mode without inducing center-of-mass motion, whereas an initially finite-momentum droplet with $p_0=\pi/2$ undergoes chirp-induced mobility arrest even without external forcing. In the presence of an external force, the linear momentum periodically modulates the lattice-dispersion contribution, but the decay of BO amplitude is governed by the nonlinear chirp response and its reweighting through the balance between the mean-field and LHY interactions. The suppression of BO is therefore not a thermodynamic damping process, but a coherent conversion of center-of-mass motion into internal phase deformation. In this sense, the LHY correction acts as a nonlinear phase-response modifier, converting an initially coherent BO into a progressively arrested trajectory.

The real-time two-level simulations further confirm the analytical picture by showing how the nonlinear corrections appear in the tunneling rate itself. For the forward process, the simulations preserve an LZ-like sweep-rate dependence, with the exponent renormalized by the nonlinear weighting factor predicted from the complex-action analysis. In contrast, the reverse transition shows a broadened, plateau-like response at stronger nonlinearity, captured by a generalized fit with a modified exponent and prefactor. This deviation implies the nonreciprocal transport. The different initial nonlinear branches carry a distinct population-imbalance background, producing an interaction-induced inertia against state variation.

Taken together, the deep- and shallow-lattice analyses show that the LHY correction does not merely stabilize the droplet as a static self-bound object. It changes the dynamical route through which lattice driving is absorbed by the system. In the tight-binding regime, this route is the chirp-mediated feedback between the center-of-mass motion and the internal phase curvature. Near the avoided crossing, it is the action and phase-space structure that control the nonlinear tunneling. The common mechanism is therefore a redistribution of driven transport into the internal nonlinear degrees of freedom, governed by the balance between the mean-field and the LHY responses.

Finally, we emphasize that the present analysis is based on a minimal two-level reduction that isolates the conservative core of the droplet dynamics. Extending beyond this reduced description by incorporating additional dynamical components may lead to effective non-Hermitian features in the reduced dynamics, reflecting the coupling to neglected degrees of freedom. Such extensions are expected to further enrich the nonadiabatic dynamics and will be explored in an upcoming work.

%
%

\appendix

\section{Lagrangian coefficient functions}
\label{App:deep-abcd}
\numberwithin{equation}{section}

In Eq.~(\ref{Eq:Lagranian}), the coefficient functions correspond to $a(m) = 2^{-1}2^{-1/m}{\Gamma(3/2m)}/{\Gamma(1/2m)}$, $b(m) = 2^{1/2m}2^{m-1}(2m\pi)^{1/2}(2m-1)^{-1/2} J/{\Gamma(1/2m)}$, $c(m) = {2}^{-1}m{A}/{\Gamma(1/2m)}$, and $d(m) = 2^{1/4m} m^{1/2}$ $\left(2/3\right)^{1+1/2m} B/{\sqrt{\Gamma(1/2m)}}$, respectively. Here, $\Gamma(x)$ is the Gamma function.

\section{Algebraic details of the quartic discriminant analysis}
\label{App:shallow-1}

\numberwithin{equation}{section}

In Sec.~IV, the steady-state population imbalance $Q$ is derived from the quartic equation:  
\begin{align}
Q^4 + \frac{2 v_T}{s_{ML}}Q^3 + \left(\frac{v_T^2}{s_{ML}^2} + \frac{V_0^2}{4 s_{ML}^2} - 1\right)Q^2 - \frac{2v_T}{s_{ML}}Q - \frac{v_T^2}{s_{ML}^2} = 0. \label{B1}
\end{align}
The boundaries of the multivalued energy band, $\pm v_c$, correspond to the emergence of four real roots to Eq.~(\ref{B1}). This occurs when the polynomial discriminant $\Delta$ becomes negative. Following the general theory of quartic equations, we define the following intermediate variables:
\begin{align}
D &= 3(2v_T/s_{ML})^2 - 8(v_T^2/s_{ML}^2 + V_0^2/4s_{ML}^2 -1) \nonumber\\
E &= -(2v_T/s_{ML})^3 + 4(2v_T/s_{ML})(v_T^2/s_{ML}^2 + V_0^2/4s_{ML}^2 -1) + 8(2v_T/s_{ML}) \nonumber\\
F &= 3(2v_T/s_{ML})^4 + 16(v_T^2/s_{ML}^2 + V_0^2/4s_{ML}^2 -1)^2 - 16(2v_T/s_{ML})^2 \nonumber\\
& \times(v_T^2/s_{ML}^2 + V_0^2/4s_{ML}^2 -1) - 16(2v_T/s_{ML})^2 
-64(v_T^2/s_{ML}^2 + V_0^2/4s_{ML}^2 -1).  \label{B2}
\end{align}
Using these, the conditions for the existence of four real roots are
$\Delta = B^2-4AC < 0$, $E \neq 0$, and $-1 < (3B-2AD)/2A\sqrt{A} < 1$, in which $A=D^2-3F$, $B=DF-9E^2$, $C=F^2-3DE^2$.

By setting $\Delta = 0$ at the transition point, the relation between $v_c$, $V_0$, and $s_{ML}$ is reduced to a sextic equation: 
\begin{align}
& v_c^6 + 3\left(\frac{V_0^2}{4} - s_{ML}^2\right)\,v_c^4 + 3\left(s_{ML}^4 + \frac{V_0^4}{16} + \frac{7}{4} s_{ML}^2 V_0^2\right)\,v_c^2 + \frac{3}{4} \left(s_{ML}^2 - \frac{V_0^2}{4}\right)\,s_{ML}^2V_0^2 \nonumber\\
& + \frac{V_0^6}{64} - s_{ML}^6 = 0. 
\end{align}
The condition for real solutions of the reduced cubic equation yields
\begin{align}
v_c^2 = \left(s_{ML}^2 - V_0^2/4 \right) + 3 s_{ML}^{2/3}(V_0/2)^{4/3} - 3 s_{ML}^{4/3}(V_0/2)^{2/3},
\end{align}
leading to the critical sweep parameter
\begin{align}
v_c = \left(s_{ML}^{2/3} - (V_0/2)^{2/3}\right)^{3/2},
\end{align}
in the regime $s_{ML} > V_0/2$.

\section{Stability Analysis and Phase-Space Linearization}
\label{App:shallow-1-phasedynamics}

\numberwithin{equation}{section}

To quantify the stability of the fixed points in the phase-space dynamics, we linearize the flow of the Josephson-analog system around each stationary solution. By introducing small perturbations, the linearized dynamics are governed by the Jacobian matrix
\begin{align}
J = \begin{bmatrix}
\frac{\partial f}{\partial Q} & \frac{\partial f}{\partial \theta} \\
\frac{\partial g}{\partial Q} & \frac{\partial g}{\partial \theta}
\end{bmatrix},
\end{align}
where $f=dQ/dt$ and $g=d\theta/dt$. The stability of each fixed point is determined by the local eigenvalues of this matrix that characterize the growth or decay of fluctuations.
In particular, the evolution near a fixed point can be expressed in terms of exponential modes $\exp(\pm \chi t)$, where the characteristic exponent $\chi$ is given by
\begin{align}
\chi = \left[ \frac{V_0^2 Q^2}{4 (1-Q^2)} 
- \frac{V_0^2 (1+Q^2)}{4(1-Q^2)}\cos^2\theta 
+ \frac{V_0}{2}s_{ML}\sqrt{1-Q^2}\cos\theta \right]^{1/2}.
\end{align}

For the hyperbolic saddle point $p_3$, $\chi$ is real and positive, leading to exponentially growing and decaying modes that define the unstable and stable manifolds. These manifolds form the separatrix structure observed in Fig.~\ref{Fig4}. 

In contrast, for elliptic fixed points, $\chi^2 < 0$, and the dynamics become oscillatory, $\exp(\pm i \chi t)$, corresponding to closed orbits in phase space. These periodic trajectories describe the self-trapped oscillations within the nonlinear regime.

The interplay between stable and unstable manifolds determines the topology of the phase space and governs the transition across the separatrix. The annihilation of the saddle point at $v_T = v_c$ marks the disappearance of the homoclinic orbit and restores the open-orbit structure at large bias.

\section{LZ Transition in the Linear Regime}
\label{App:shallow-2-2latticerule}

In this Appendix, we present a detailed derivation of the nonadiabatic transition probability for the linearized two-level system, retaining the full analytical structure underlying the exponential LZ formula.

We begin by reorganizing the linear part of the coupled two-level equations and introducing the transformation
\begin{align}
z = \alpha_L^{1/2} e^{-i\pi/4} t, \qquad 
n = \frac{i V_0^2}{16 \alpha_L} \equiv i\gamma,
\end{align}
which maps the time-dependent problem into the Weber form. The resulting equations for the amplitudes $c_a$ and $c_b$ read
\begin{align}
\frac{d^2 c_a}{dz^2} - \left(a_1 + \frac{z^2}{4}\right)c_a = 0, \qquad a_1 = -n + \frac{1}{2},
\end{align}
\begin{align}
\frac{d^2 c_b}{dz^2} - \left(a_2 + \frac{z^2}{4}\right)c_b = 0, \qquad a_2 = -n - \frac{1}{2}.
\end{align}

These equations admit solutions in terms of parabolic cylinder functions \cite{Abramowitz1965}. In particular, the solution for $c_b$ can be written as
\begin{align}
c_b(z) = A\, U(a, -iz), \qquad a = n + \frac{1}{2},
\end{align}
where $A$ is determined by the initial condition and the asymptotic behavior of $U$. In this limit $t \to -\infty$, the argument of the function becomes $-iz = e^{i\pi/4} \mathcal{R}$ with $\mathcal{R} \to \infty$. Along this direction, the asymptotic function yields
\begin{align}
c_b(z) \sim A\, e^{-i\mathcal{R}^2/4} \mathcal{R}^{-n-1} e^{-i\pi(n+1)/4},
\end{align}
which implies $|c_b| \to 0$ as required for the system to occupy the lower diabatic state.

The normalization condition $|c_a(-\infty)|^2 = 1$ then imposes a constraint on $A$, given by $\lim_{t \rightarrow -\infty} ({4}/{V_0}) (|d c_b/dt|) + |({4}/{V_0}) (\alpha_L t /2) c_b| = 1$. Evaluating the dominant contributions to the normalization in this limit yields
\begin{align}
|A| = \gamma^{1/2} e^{-\pi \gamma /4}.
\end{align}
The asymptotic behavior at $t \to +\infty$ is also determined, where the argument rotates to $-iz = \mathcal{R} e^{-i3\pi/4}$. To determine the asymptotic form in this new direction, we use the identity
\begin{align}
\sqrt{2\pi} U(a, \pm x) 
= \Gamma\!\left(\frac{1}{2} - a\right)e^{-i\pi(a/2+1/4)} U(-a, \pm ix)
+ \Gamma\!\left(\frac{1}{2} - a\right)e^{i\pi(a/2+1/4)} U(-a, \mp ix),
\end{align}
which allows the analytic continuation of the solution.

Applying this relation and retaining the leading contributions for large $\mathcal{R}$, we obtain
\begin{align}
c_b(z) \sim 
\frac{\sqrt{2\pi}}{\Gamma(n+1)} \gamma^{1/2} e^{-\pi\gamma/2} e^{i \mathcal{R}^2/4} e^{i\gamma}.
\end{align}
From this expression, the final occupation probability follows as
\begin{align}
|c_b(\infty)|^2 
= \frac{2\pi \gamma}{\Gamma(1+i\gamma)\Gamma(1-i\gamma)} e^{-\pi\gamma}
= 2 e^{-\pi\gamma} \sinh(\pi\gamma).
\end{align}
The nonadiabatic transition probability is therefore given by
\begin{align}
\Gamma_{\mathrm{LZ}} = 1 - |c_b(\infty)|^2 = e^{-2\pi\gamma}.
\end{align}
Substituting $\gamma = V_0^2 / 16\alpha_L$, we recover the standard LZ expression
\begin{align}
\Gamma_{\mathrm{LZ}} = \exp\!\left(-\frac{\pi V_0^2}{8\alpha_L}\right),
\end{align}
which characterizes the exponential suppression of tunneling in the adiabatic limit.

%
%

%
%
%
\begin{acknowledgements}
     We thank the Ministry of Science and Technology, Taiwan for partial financial support under grants NSTC 115-2918-I-845-003, NSTC 114-2221-E-845-002 and NSTC 112-2112-M-034-001. W.H.K thanks for the hospitality of the Science Institute of the University of Iceland. 
     
\end{acknowledgements}
%
%
%
\bibliographystyle{andp2012.bst}
\providecommand{\WileyBibTextsc}{}
\let\textsc\WileyBibTextsc
\providecommand{\othercit}{}
\providecommand{\jr}[1]{#1}
\providecommand{\etal}{~et~al.}

%
%
%
\end{document}